\documentclass[namedreferences]{solarphysics}
\usepackage[optionalrh]{spr-sola-addons}
\usepackage{epsfig}
\usepackage{graphicx}
\usepackage{color}
\usepackage{url}
\usepackage{lscape}
\usepackage{sidecap}
\newcommand{\etal}{{\it et al.}}
\usepackage[pdfborder={0 0 0 },urlcolor=blue]{hyperref}
\ifx \doiurl \undefined \def \doiurl#1{\href{http://dx.doi.org/#1}{\url{#1}}}\fi
\ifx \adsurl \undefined \def \adsurl#1{\href{http://adsabs.harvard.edu/abs/#1}{\url{#1}}}\fi

\begin{document}
\begin{article}
\begin{opening}

\title{Long--Term Variations in Solar Differential Rotation and Sunspot Activity, II: Differential Rotation Around the Maxima and Minima of Solar Cycles~12\,--\,24}  

\author{J. Javaraiah$^*$}

\institute{Bikasipura, Bengaluru-560 111,  India.\\
$^*$(Formerly working at Indian Institute of Astrophysics, Bengaluru-560 034, India.)\\
email: \url{jajj55@yahoo.co.in;  jdotjavaraiah@gmail.com}\\
}
\runningauthor{J. Javaraiah}
\runningtitle{Long--Term Variation in Solar Differential Rotation}

\begin{abstract}
Studies of variations in the solar differential rotation are important for
understanding the underlying mechanism of solar cycle and other variations 
of solar activity. 
We analyzed the  sunspot-group daily data that were  reported by Greenwich
 Photoheliogrphic Results (GPR) during the period 1874\,--\,1976
 and Debrecen  Photoheliographic Data (DPD) during the period 1977\,--\,2017.
 We determined  the equatorial rotation rate [$A$] and the latitude
 gradient [$B$] components of the  solar differential rotation by fitting 
 the data in each of the 3-year moving time 
intervals (3-year MTIs) successively shifted by one year during the
 period 1874\,--\,2017 
to the standard law of differential rotation. 
The values of $A$ and $B$ around the years of maxima and
minima of  Solar Cycles 12\,--\,24   are obtained from the 3-year MTIs
series of $A$ and $B$ and studied
the long--term cycle-to-cycle modulations in these coefficients.
Here we have used the epochs of the maxima and  minima
 of  the  Solar Cycles 12\,--\,24 that
 were recently determined from the revised Version-2 international 
sunspot-number series. We  find that
there exits  a considerably significant secular decreasing-trend in
  $A$  around the maxima of solar cycles. There exist
 no secular trends in both  $A$ and $B$   around the minima of solar
cycles. The secular trend in $B$   around the  maxima
 of solar cycles is also found to be statistically insignificant.
  We fitted a cosine function to the values of $A$, and also
to those  of $B$, after removing the corresponding linear trends. 
The cosine-fits suggest
that there exist $\approx$54-year ($\approx$94-year) and 
$\approx$82-year ($\approx$79-year) periodicities in
$A$ ($B$)  around the maxima  and minima of solar cycles, respectively. 
The amplitude  of the cosine-profile of
 $A$ ($B$)  around the minima of solar cycles 
 is  about 41\,\% (65\,\%) larger than that 
 of  $A$ ($B$) of  around the  maxima. 
 In addition, the cosine profiles of $A$ and $B$ suggest  a 
large (up to $180^\circ$) phase difference between  the long-term variations
 of $A$, and also  between those of $B$, of  around  maxima and 
minima of solar cycles.
Implications of all these results are  discussed briefly.
\end{abstract}

\end{opening}

\section{Introduction}
Solar dynamo models explain several properties of solar cycle
 (\opencite{bab61}; \opencite{dg06}; \opencite{camer17}). 
The solar differential rotation is an important ingredient in these models.
Solar activity vary in many time scales. 
Therefore, studies of variations in solar differential rotation 
are important  for better understanding the physical mechanism of 
 solar cycle and other variations of solar activity. 
Variations in the differential rotation of the Sun's outer layers have been 
known for several decades. 
Many authors studied the solar cycle  variation and other short- and long--term 
variations  in the  differential rotation by using 
 different data and techniques
(\opencite{hr76}; \opencite{gh84}; \opencite{sh85}; \opencite{jsb85};
 \opencite{balth86}; \opencite{kn90}; \opencite{snod92}; \opencite{khh93}; 
  \opencite{yk93}; \opencite{jg95}; \opencite{mak97}; \opencite{jj03}, 
 \citeyear{jj05}, \citeyear{jj11}, \citeyear{jj13}; \opencite{gupta99}; 
\opencite{jk99}; \opencite{kgk02}; \opencite{braj06}; \opencite{ju06}; 
 \opencite{svand08}; \opencite{jub09}; \opencite{cvi10}; \opencite{li14},
 \opencite{jb16}; \opencite{os16}; \opencite{bo17}; \opencite{ruz17}; 
 \opencite{roudi18}). 
However, so far only the 11-year period torsional oscillation of photospheric 
layers was discovered by  \inlinecite{hl80} and \inlinecite{lh82} is  
established.  
Helioseismic techniques have revealed that the torsional oscillation pattern
extends inward up to about one-third  thickness of the 
  Sun's convection zone~(\opencite{howe00}). 
Other results to some extent established  are: both equatorial rate  and 
latitude gradient of  rotation are  large at solar minimum than at 
maximum, and we have a secular decrease of the solar equatorial rotation 
with  secular 
increase of activity (see \opencite{ruz17} and the references therein).  

\inlinecite{jbu05a} (hereafter Article-I)  analyzed the Greenwich and 
 Solar Optical Observing Network (SOON)
 sunspot-group data during the period 1874\,--\,2007 and determined the 
average values of the coefficients of differential rotation during each of the 
 Solar Cycles 12\,--\,22 to study  the cycle-to-cycle 
modulation in these coefficients  and found the secular decrease 
(about $-0.01^\circ$ day$^{-1}$ cycle$^{-1}$) of  
the cycle-to-cycle modulation in the   equatorial rotation rate.
\inlinecite{balth86}, \inlinecite{braj06},
 and \inlinecite{li14} have also studied in 
detail the cycle-to-cycle  variations in the solar 
differential rotation. 
  In Article-I 
cosine functions to the  values of the coefficients of  differential 
rotation were fitted and we  found  the existence of a 79-year cycle
 (Gleissberg cycle) in the latitude gradient of rotation. 
 \inlinecite{suz14} found that  a period of 
about 6--7 solar cycles exits in  long-term modulation of the
 latitude gradient of differential rotation. It should be noted here 
a number of authors have found  changes in  the period (60--130-year) 
of the Gleissberg
cycle (\opencite{garc98}; \opencite{hath99}; \opencite{roze94}; 
\opencite{ogurt02}; \opencite{hath15}; \opencite{jj17}).
Recently, \inlinecite{jj19} determined  cycle-to-cycle 
variations in the sunspot activity at the minima and the maxima of 
 Solar Cycles 12\,--\,24 and  noticed  that there exists a
 difference in the long-term periodicities of  the activity at
 solar minimum and maximum. 
The rotation rates of 
magnetic active regions (sunspot groups, etc.) vary with  life 
times and sizes (areas) of the active regions. 
Large/long-lived sunspot groups
rotate more slowely than small/short-lived sunspot groups
(\opencite{war65}, \citeyear{war66}; \opencite{hgg84}; \opencite{balth86}). 
 This could have to do with the magnetic structures of large and 
small sunspot groups anchoring at deep and shallow layers of the 
solar convection
 zone, respectively~(\opencite{jg97}; \opencite{hi02}; \opencite{siva03}).
  The average size of sunspot groups in the minimum of a solar cycle 
 is smaller than that of the sunspot groups in the maximum. Hence, the study 
of solar differential rotation  of the sunspot groups at solar minimum and 
maximum may be useful for better understanding the physical  mechanism of
the solar long-term variability. In the present analysis we determined 
the values of the coefficients of differential rotation in 
the 3-year MTIs during 
the period 1874\,--\,2017. From  the obtained time series     
we determined  the 
values of the coefficients of differential rotation of the sunspot groups 
 around the minima and  maxima of  Solar Cycles 12\,--\,24 and 
separately study 
the long-term cycle-to-cycle modulations in the coefficients  
of the differential rotation  around the maxima and minima of solar cycles.      

In the next section we describe the data and method of analysis. 
In Section~2 we present  results and in Section~3 we present conclusions 
and discuss upon them briefly.

\section{Data and Method of Analysis}
Here we have used the  daily sunspot-group data reported 
 in GPR during the period
 April 1874\,--\,December 1976
 and  DPD during the period
 January 1977\,--\,June 2017.
These data are downloaded from the website 
{\sf fenyi.\break solarobs.unideb.hu/pub/DPD/}.
The  details about these data can be found in  
  \inlinecite{gyr10}, \inlinecite{bara16}, and \inlinecite{gyr17}. 
 These data contain the date and time  of observation (we converted these into 
 date with fractional day, [$t$]), heliographic 
latitude [$\lambda$] and longitude [$L$], corrected whole-spot area, 
and central 
meridian distance [CMD], etc. of a sunspot group for each of 
the days it was observed.  
Here each disk passage of a recurrent sunspot 
group is treated as an independent sunspot group and 
in order to reduce the foreshortening effect (if any), we 
have not used the data in any  day of the sunspot 
group life-time/disk passage ([$\tau$], its values are 2, 3,$\dots$,12 days).  
 We have determined the 
 equatorial 
rotation rate [$A$] and the latitudinal gradient [$B$] of the solar
differential  rotation by fitting the 
data corresponding to  a specified time-interval to  the standard law 
of differential rotation: 
$$ \omega (\theta) = A + B \sin^2(\theta),  \eqno(1)$$ 
where $\omega(\theta) = \frac{L_i-L_{i -1}}{t_i - t_{i-1}} + 14^\circ.18$ is
 the solar sidereal rotation rate at the latitude 
$\theta = \lambda_{i-1}$,  $i =$ 2, 3,$\dots$,$\tau$, and 
 $14^\circ.18$ is the Carrington rigid body rotation rate of the Sun. 
In all our earlier analyzes we have 
assigned the velocity value to the mean of $\lambda_{i-1}$  and $\lambda_i$.
Following the suggestion by  \inlinecite{ok05} here we assigned the velocity 
value to the $\lambda_{i-1}$ (also see \opencite{sudar14}).  
 The data correspond to  $\omega > 3^\circ$ 
 day$^{-1}$ and the latitudinal drift 
$\frac{\lambda_i-\lambda_{i -1}}{t_i - t_{i-1}}> 2^\circ$ day$^{-1}$ 
are excluded.  This considerably reduced the 
uncertainties (standard deviations) in the derived values of $A$ and 
$B$~(\opencite{war65}, \citeyear{war66}; \opencite{jg95}). The data 
corresponding  $t_i - t_{i-1} >2$ days
 (non-consecutive days) are also excluded. This further reduced the 
uncertainties in $A$ and $B$. In order to have 
a better statistics (particularly during solar minimum) first
 we have fitted  Equation~1 to the data in the  
3-year MTIs
 1874\,--\,1876, 1875\,--\,1877,$\dots$,2015\,--\,2017
during the period 1874\,--\,2017.  Northern- and southern-hemispheres'
data are combined. 
  The values of $A$ and $B$ around the years of maxima and 
minima ($i.e.$ in the 3-year intervals in which 
the years of maxima and minima are middle years) of the
 Solar Cycles 12\,--\,24   are obtained from the 3-year MTIs 
series of $A$ and $B$. We have used the epochs of the maxima and the minima,
cycle-lengths, $etc.$ of  the  Solar Cycles 12\,--\,24 that
 were recently determined   
by~\inlinecite{pesnell18} using the revised Version-2 international 
sunspot-number (SN) series.

\section{Results}
 In Table~1, we have given the values of  mean
 sidereal angular velocity ($\omega$, in degree day$^{-1}$)
 and  corresponding uncertainty (standard error, $\delta \omega$)
 in  different $2^\circ$ latitude ($\theta$) intervals, 
 determined from the sunspot-group data around the minima
($i.e.$ from the combined data in the intervals 1877--1879, 1889--1891,
 1901--1903, 1912--1914, 1922--1924, 1932--1934, 1943--1945,
 1953--1955, 1975--1977, 1985--1987, 1995--1997, and 2007--2009)
and around the maxima (i.e. from the combined data in the intervals
1882--1884, 1893--1895, 1905--1907, 1916--1918, 1927--1929, 1936--1938,
1946--1948, 1957--1959, 1967--1969, 1978--1980, 1988--1990,
2000--2002, and 2013--2015) of Solar Cycles 12--24. In this table 
 the corresponding number ($N$) of the
velocity values in each latitude interval is also given. Northern- and 
southern-hemispheres' data are folded. In many latitude intervals
 $N$ is considerably large.  Hence,  the corresponding 
uncertainties in the values of the mean angular velocities are 
reasonably small. 

\begin{table}
\caption{Values of  mean ($\omega$, in degree day$^{-1}$) 
  and  corresponding standard error  ($\delta \omega$) 
 of sidereal angular velocity 
 in  different $2^\circ$ latitude ($\theta$) intervals 
 determined from the sunspot-group data around the minima
($i.e.$ from the combined data in the intervals: 1877--1879, 1889--1891, 
 1901--1903, 1912--1914, 1922--1924, 1932--1934, 1943--1945, 
 1953--1955, 1975--1977, 1985--1987, 1995--1997, and 2007--2009), 
and around the maxima (i.e. from the combined data in the intervals: 
1882--1884, 1893--1895, 1905--1907, 1916--1918, 1927--1929, 1936--1938, 
1946--1948, 1957--1959, 1967--1969, 1978--1980, 1988--1990, 
2000--2002, and 2013--2015) of Solar Cycles 12--24.
 The corresponding number ($N$) of the 
velocity values in each latitude interval is also given. Northern- and 
southern-hemispheres' data are folded.}

\begin{tabular}{lccccccc}
\noalign{\smallskip}
\hline
Latitude & \multicolumn{3}{c}{Around minimum} && \multicolumn{3}{c}{Around maximum}\\
\cline{2-4}
\cline{6-8}
interval &$\omega$ & $\delta \omega$ &$N$&&$\omega$
&$\delta \omega$&$N$\\
\hline
$0^\circ$--$2^\circ$& 14.57&  0.05&   318&& 14.56&  0.04&   550\\
$2^\circ$--$4^\circ$& 14.50&  0.04&   473&& 14.56&  0.02&  1452\\
$4^\circ$--$6^\circ$& 14.56&  0.03&   830&& 14.52&  0.02&  3393\\
$6^\circ$--$8^\circ$& 14.46&  0.03&   875&& 14.46&  0.01&  5412\\
$8^\circ$--$10^\circ$& 14.47&  0.03&  1094&& 14.45&  0.01&  7627\\
$10^\circ$--$12^\circ$& 14.39&  0.03&   805&& 14.40&  0.01&  9350\\
$12^\circ$--$14^\circ$& 14.44&  0.04&   542&& 14.38&  0.01& 10290\\
$14^\circ$--$16^\circ$& 14.38&  0.04&   642&& 14.33&  0.01& 10136\\
$16^\circ$--$18^\circ$& 14.30&  0.04&   608&& 14.28&  0.01&  9575\\
$18^\circ$--$20^\circ$& 14.28&  0.03&   866&& 14.25&  0.01&  8459\\
$20^\circ$--$22^\circ$& 14.21&  0.03&  1147&& 14.18&  0.01&  6528\\
$22^\circ$--$24^\circ$& 14.22&  0.03&  1081&& 14.11&  0.01&  5086\\
$24^\circ$--$26^\circ$& 14.09&  0.04&   796&& 14.05&  0.02&  3914\\
$26^\circ$--$28^\circ$& 14.02&  0.04&   700&& 13.96&  0.02&  2681\\
$28^\circ$--$30^\circ$& 14.06&  0.05&   520&& 13.90&  0.02&  1792\\
$30^\circ$--$32^\circ$& 13.75&  0.05&   357&& 13.86&  0.03&  1224\\
$32^\circ$--$34^\circ$& 13.67&  0.07&   212&& 13.74&  0.04&   811\\
$34^\circ$--$36^\circ$& 13.88&  0.12&   104&& 13.69&  0.05&   492\\
$36^\circ$--$38^\circ$& 13.68&  0.14&    59&& 13.71&  0.06&   321\\
$38^\circ$--$40^\circ$& 13.42&  0.17&    37&& 13.39&  0.09&   154\\
\hline
\end{tabular}
\label{table1}
\end{table}

Figure~1 shows the mean latitude dependence of the solar sidereal rotation rate
 [$\omega (\theta)$] around the minima and around the maxima of Solar
 Cycles 12\,--\,24 determined from the sunspot-group data of these cycles.
 In this figure we have shown the  profiles of the differential rotation
 deduced from the values of the coefficients $A$ and $B$ of Equation~1, and
 also  the values of $\omega(\theta)$  determined
 by averaging the  data into $2^\circ$ latitude intervals,  which are also  
given in Table~1. Northern- and southern-hemispheres' data are folded.
As can be seen in this figure, as already known ($e.g.$
\opencite{balth86}; \opencite{khh93}; \opencite{gupta99}; \opencite{jj03};
\opencite{braj06}; \opencite{ruz17}),  the mean angular velocity of 
the sunspot groups  around the minima of the solar cycles is slightly
 larger than that of the sunspot groups  around the maxima in all
 latitudes (see the dashed and the continuous curves).  Around the 
minima of the solar cycles the mean  equatorial velocity $A$  is about
 0.16\,\% larger than that   around the maxima (this difference is
 statistically significant on at least 1.9$\sigma$ level, 
where $\sigma$ is standard deviation of $A$  around minima). 
The mean latitude gradient $B$  around the maxima is about 4.4\,\% larger
 than that  around the 
minima (it is consistent with the fact that dynamo action is strong during 
maximum and weak during minimum), but this difference is statistically highly 
insignificant, i.e. only 1.2$\sigma$ level (around the minima the
 number of data points is about 86\,\% 
lower than that of the around the maxima). On the other hand, there exist 
considerable cycle-to-cycle variations in both $A$ and $B$  around
  the maxima as well as those  around the minima of 
Solar Cycles 12\,--\,24 (see below).  

\begin{SCfigure}
\setcounter{figure}{0}
\centering
\vspace{1.9cm}
\includegraphics[width=6.5cm]{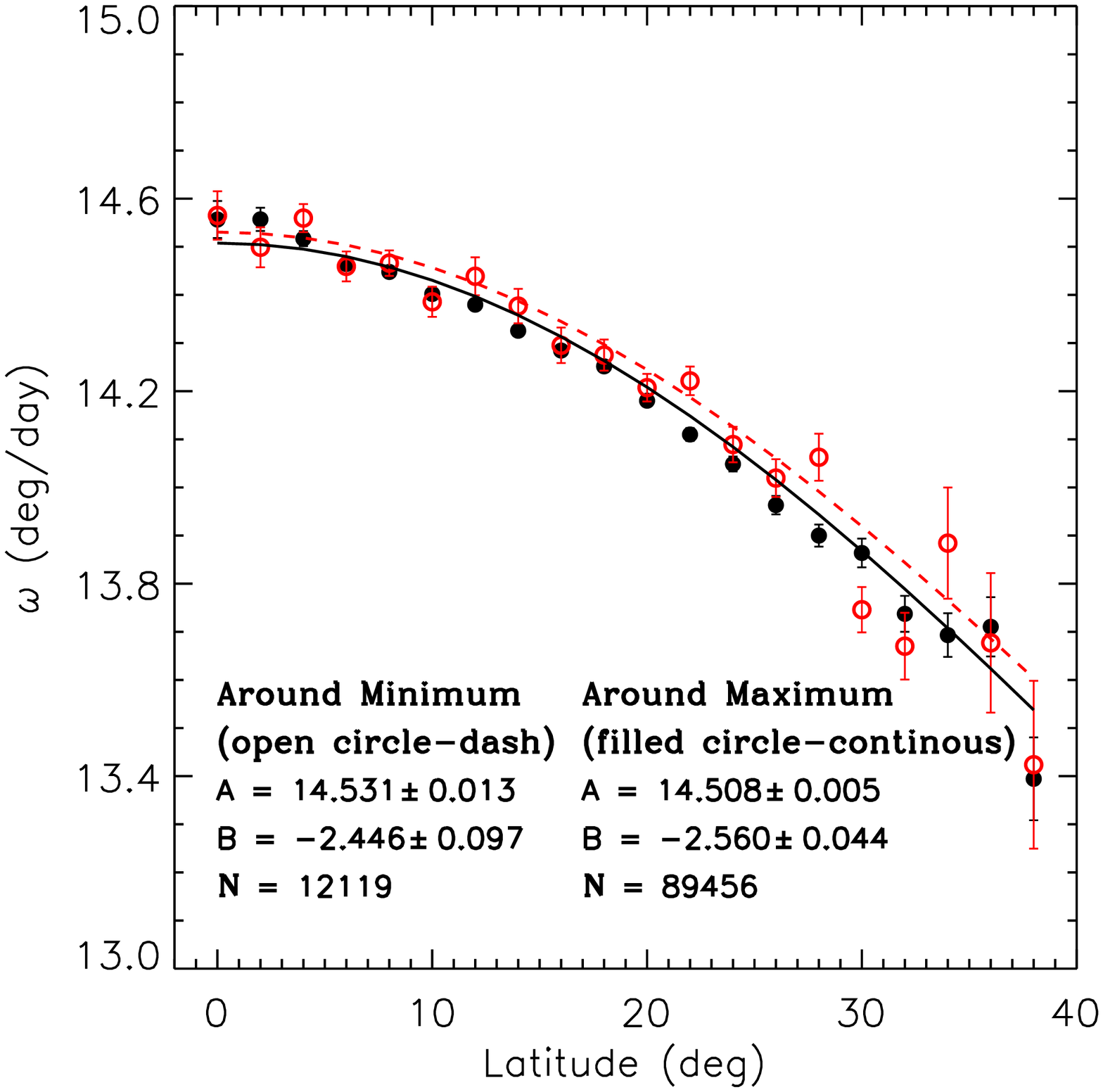}
{\tiny
\caption{The mean latitude dependence of the solar sidereal rotation rate
 [$\omega(\theta)$]  around the minima and  around the maxima
 ($i.e.$ in the 3-year intervals in which
the epochs of minima and maxima are  middle years)   
of  Solar Cycles 12\,--\,24.
The {\it continuous curve} (black)  and {\it dashed curve} (red) show the 
corresponding profiles of the differential rotation deduced from the values
 of the coefficients $A$ and $B$ in Equation~1.
 These values  of the coefficients and the number of velocity values ($N$)
 gone in the calculations  are also shown.  Northern- and 
southern-hemispheres' data are combined.  The {\it filled circles} (black)
 and {\it open circles} (red) represent the values of
 $\omega(\theta)$ 
that are  given in Table~1 and determined by averaging the  data 
into $2^\circ$ latitude intervals.} }
\label{f1}
\end{SCfigure}

\begin{figure}
\setcounter{figure}{1}
\centering
\includegraphics[width=12cm]{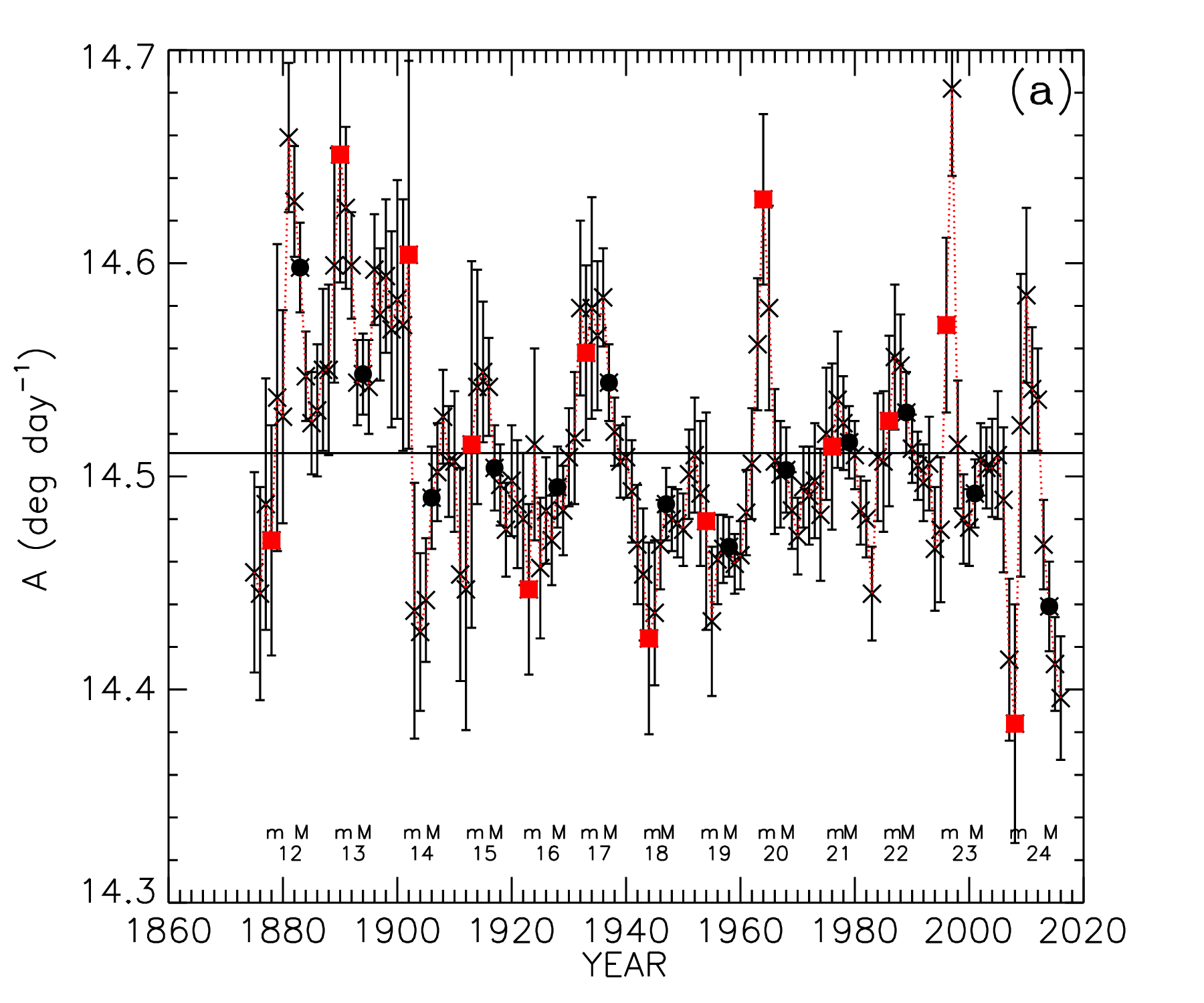}
\includegraphics[width=12cm]{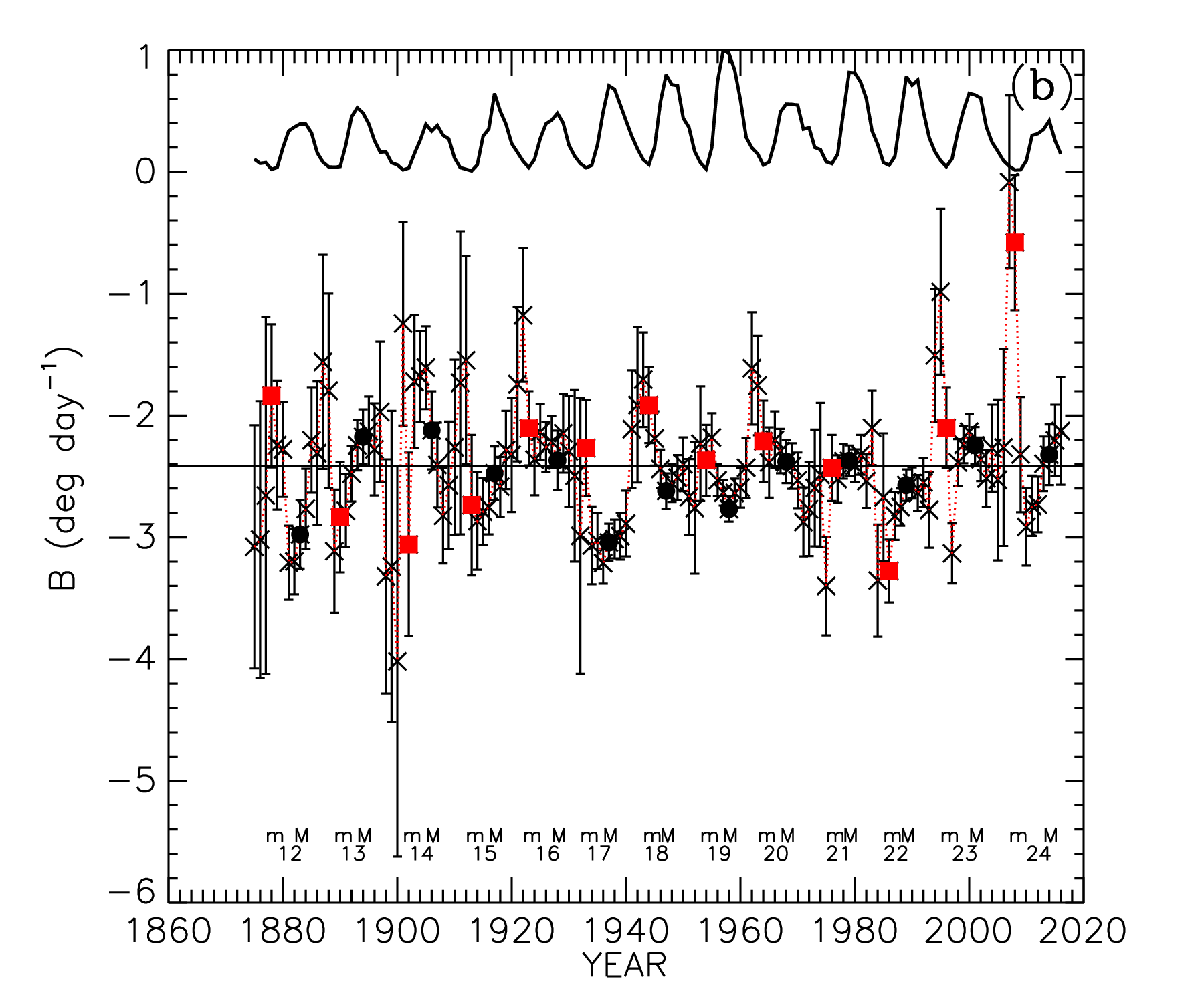}
\caption{The values ({\it open circles} conected by {\it dotted curve})
 of $A$ and $B$ determined from the
 sunspot group 
data in the 3-year MTIs 1874\,--\,1876, 1875\,--\,1877,$\dots$,2015\,--\,2017 
during the period 1874\,--\,2017 {\it versus} the middle years 1975, 
1876,$\dots$,2016 of the intervals.  Northern- and southern-hemispheres'
data are combined.  
The values at the epochs (middle years
 of the corresponding 3-year intervals) of minimum  and maximum  are
indicated with the {\it filled-square} and  {\it filled-circle}, 
respectively. The Waldmeier numbers of the solar 
cycles and the corresponding epochs of the minimum and maximum are also
 shown by the symbols m and M, respectively. 
The {\it horizontal continuous-line} represents the mean value over the 
whole period 
1874\,--\,2017 (mean $A = 14^\circ.51$ day$^{-1}$ and mean 
$B = -2^\circ.42$ day$^{-1}$). In {\bf b} the continuous curve represents 
the variation in yearly mean total sunspot number during 1875--2016.}   
\label{f2}
\end{figure}

\begin{table}
{\scriptsize
\caption[]{Values of $A$ and $B$ (in degree day$^{-1}$) 
in the 3-year intervals at the epochs  $T_{\rm m}$ and $T_{\rm M}$
 (middle years of the corresponding 3-year intervals) of the minima 
and maxima (indicated by the suffixes  m and M, respectively)
of Solar Cycles 12\,--\,24. The corresponding  number ($N$) of 
velocity values is given. Northern- and southern-hemispheres'
data are combined.
 The  minimum ($R_{\rm m}$) and maximum 
($R_{\rm M}$) Version-2 sunspot numbers are also given.  $^a$ implies
tentative (see text).}
\begin{tabular}{lcccccccccccc}
\hline
Cycle&$R_{\rm m}$& Interval &$T_{\rm m}$ & $N$ & $A_{\rm m}$ & $B_{\rm m}$\\ 
   12&3.7& 1877--1879& 1878&  314& $14.470 \pm 0.054$&$ -1.838 \pm  0.588$\\
   13&8.3& 1889--1891& 1890& 1188& $14.651 \pm  0.060$&$ -2.834 \pm 0.454$\\
   14&4.5& 1901--1903& 1902&  654& $14.604 \pm  0.091$&$ -3.058 \pm  0.754$\\
   15&2.5& 1912--1914& 1913&  429& $14.515 \pm  0.086$&$ -2.736 \pm  0.577$\\
   16&9.4& 1922--1924& 1923&  928& $14.447 \pm 0.040$&$ -2.105 \pm  0.305$\\
   17&5.8& 1932--1934& 1933&  680& $14.558 \pm  0.041$&$ -2.267 \pm  0.394$\\
   18&12.9& 1943--1945& 1944& 1346& $14.424 \pm  0.045$&$ -1.915 \pm  0.311$\\
   19&5.1& 1953--1955& 1954& 1221& $14.479 \pm  0.051$&$ -2.369 \pm  0.292$\\
   20&14.3& 1963--1965& 1964& 1135& $14.630 \pm  0.040$&$ -2.211 \pm  0.334$\\
   21&17.8& 1975--1977& 1976& 1288& $14.514 \pm  0.039$&$ -2.431 \pm  0.272$\\
   22&13.5& 1985--1987& 1986& 1215& $14.526 \pm  0.040$&$ -3.277 \pm  0.259$\\
   23&11.2& 1995--1997& 1996& 1204& $14.571 \pm  0.041$&$ -2.102 \pm  0.330$\\
   24&2.2& 2007--2009& 2008&  517& $14.384 \pm  0.056$&$ -0.580 \pm  0.555$\\
& & 2008--2010& 2009$^a$&  784& $14.524 \pm  0.071$&$ -2.321 \pm  0.4734$\\    
\\

 & $R_{\rm M}$&&$T_{\rm M}$ &  &$A_{\rm M}$& $B_{\rm M}$\\
   12&124.4& 1882--1884& 1883& 4043& $14.598 \pm  0.021$& $-2.976 \pm  0.281$\\
   13&146.5& 1893--1895& 1894& 5694& $14.548 \pm  0.019$& $-2.175 \pm  0.226$\\
   14&107.1& 1905--1907& 1906& 4037& $14.490 \pm  0.024$& $-2.122 \pm  0.321$\\
   15&175.7& 1916--1918& 1917& 5998& $14.504 \pm  0.020$& $-2.475 \pm  0.221$\\
   16&130.2& 1927--1929& 1928& 5171& $14.495 \pm  0.019$& $-2.368 \pm  0.245$\\
   17&198.6& 1936--1938& 1937& 6870& $14.544 \pm  0.018$& $-3.042 \pm  0.154$\\
   18&218.7& 1946--1948& 1947& 8006& $14.487 \pm  0.017$& $-2.620 \pm  0.144$\\
   19&285.0& 1957--1959& 1958& 10379& $14.467 \pm  0.014$& $-2.767 \pm  0.104$\\
   20&156.6& 1967--1969& 1968& 6367& $14.503 \pm  0.020$& $-2.379 \pm  0.177$\\
   21&232.9& 1978--1980& 1979& 9186& $14.516 \pm  0.017$& $-2.373 \pm  0.127$\\
   22&212.5& 1988--1990& 1989& 7853& $14.530 \pm  0.019$& $-2.572 \pm  0.129$\\
   23&180.3& 2000--2002& 2001& 8849& $14.492 \pm  0.016$& $-2.245 \pm  0.151$\\
   24&116.4& 2013--2015& 2014& 7003& $14.439 \pm  0.021$& $-2.322 \pm  0.250$\\
\hline
\end{tabular}
\label{table2}
}
\end{table}

\begin{figure}
\setcounter{figure}{0}
\centering
\begin{subfigure}
\setcounter{figure}{0}
\includegraphics[width=5.6cm]{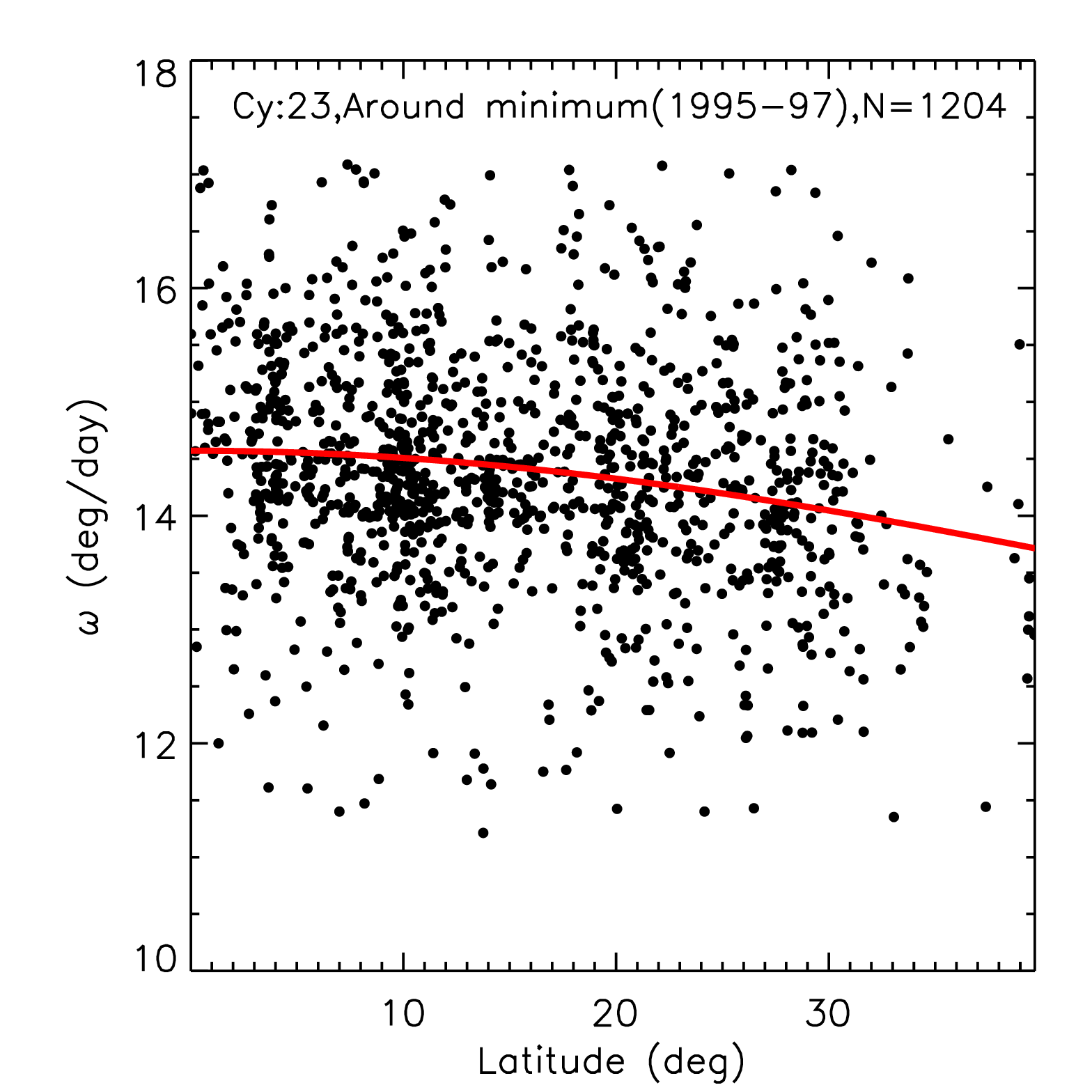}
\includegraphics[width=5.6cm]{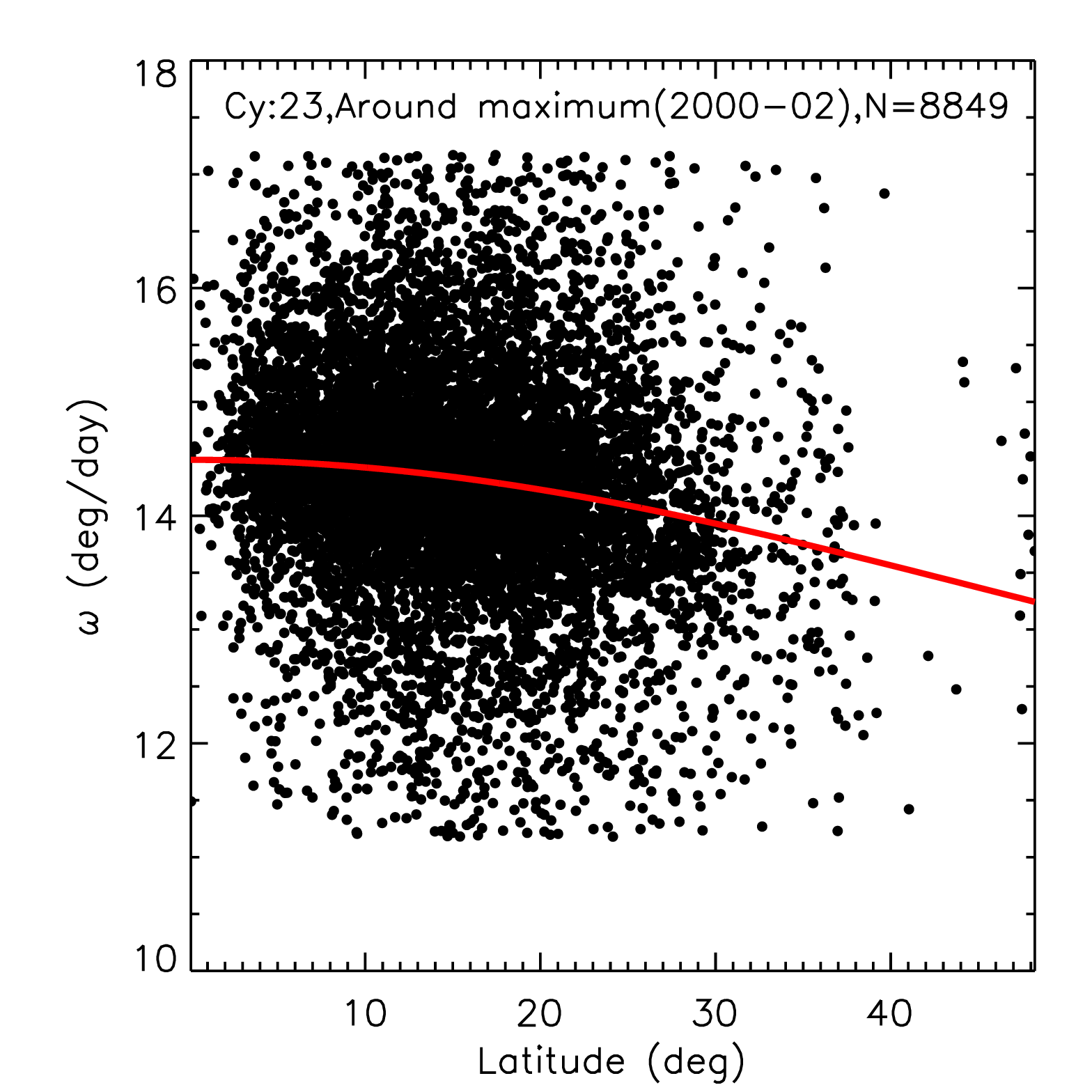}
\end{subfigure}
\begin{subfigure}
\setcounter{figure}{1}
\includegraphics[width=5.6cm]{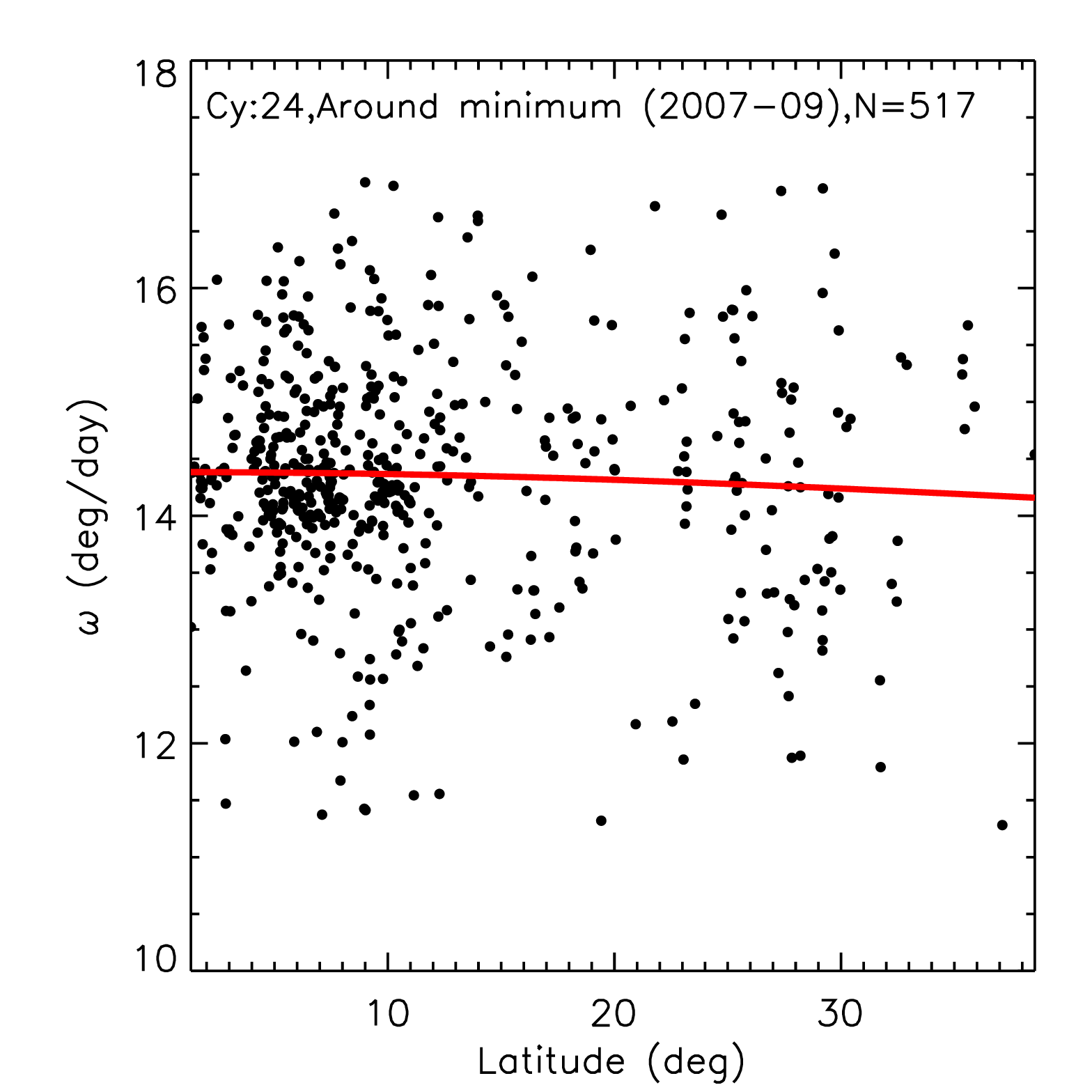}
\includegraphics[width=5.6cm]{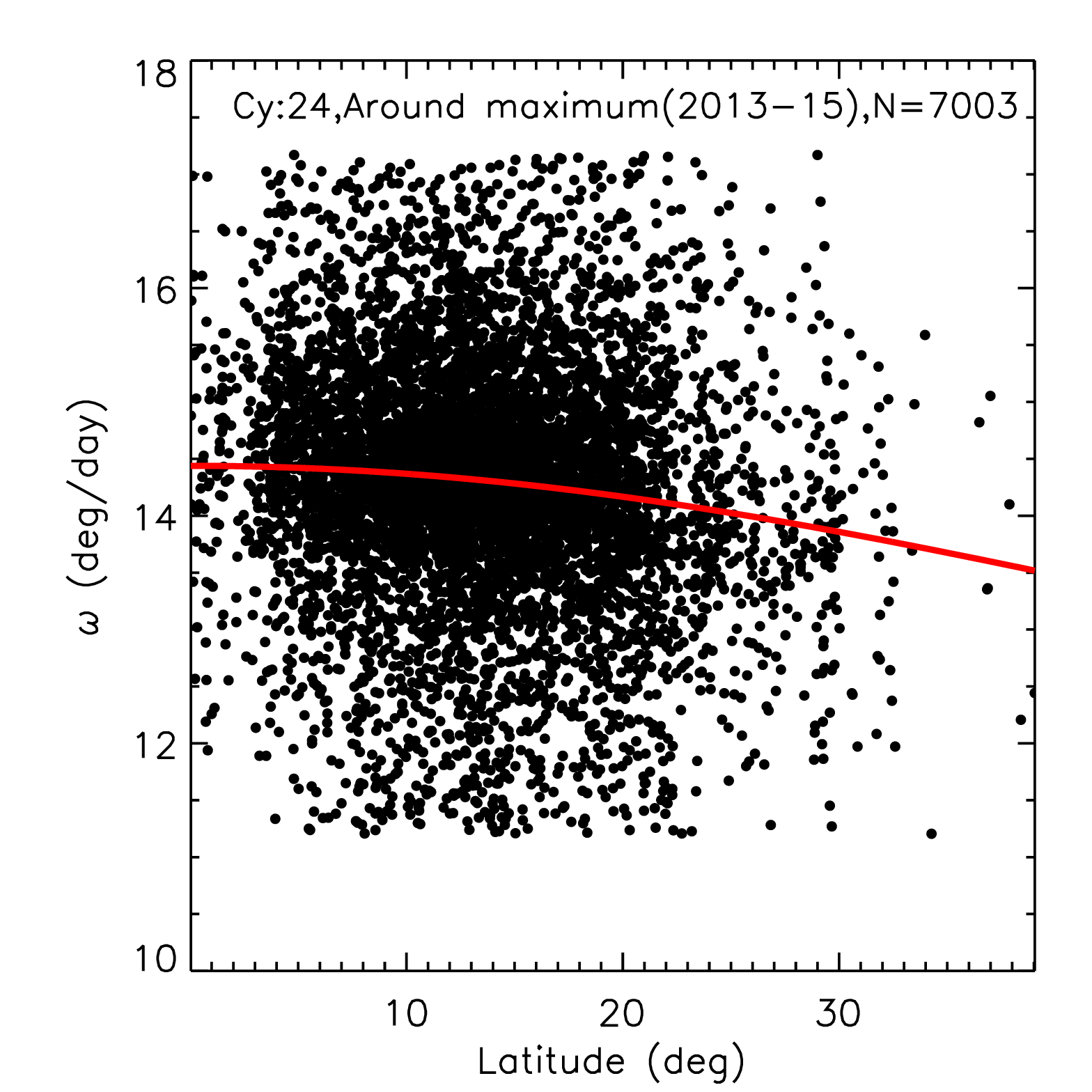}
\end{subfigure}
\begin{subfigure}
\includegraphics[width=5.6cm]{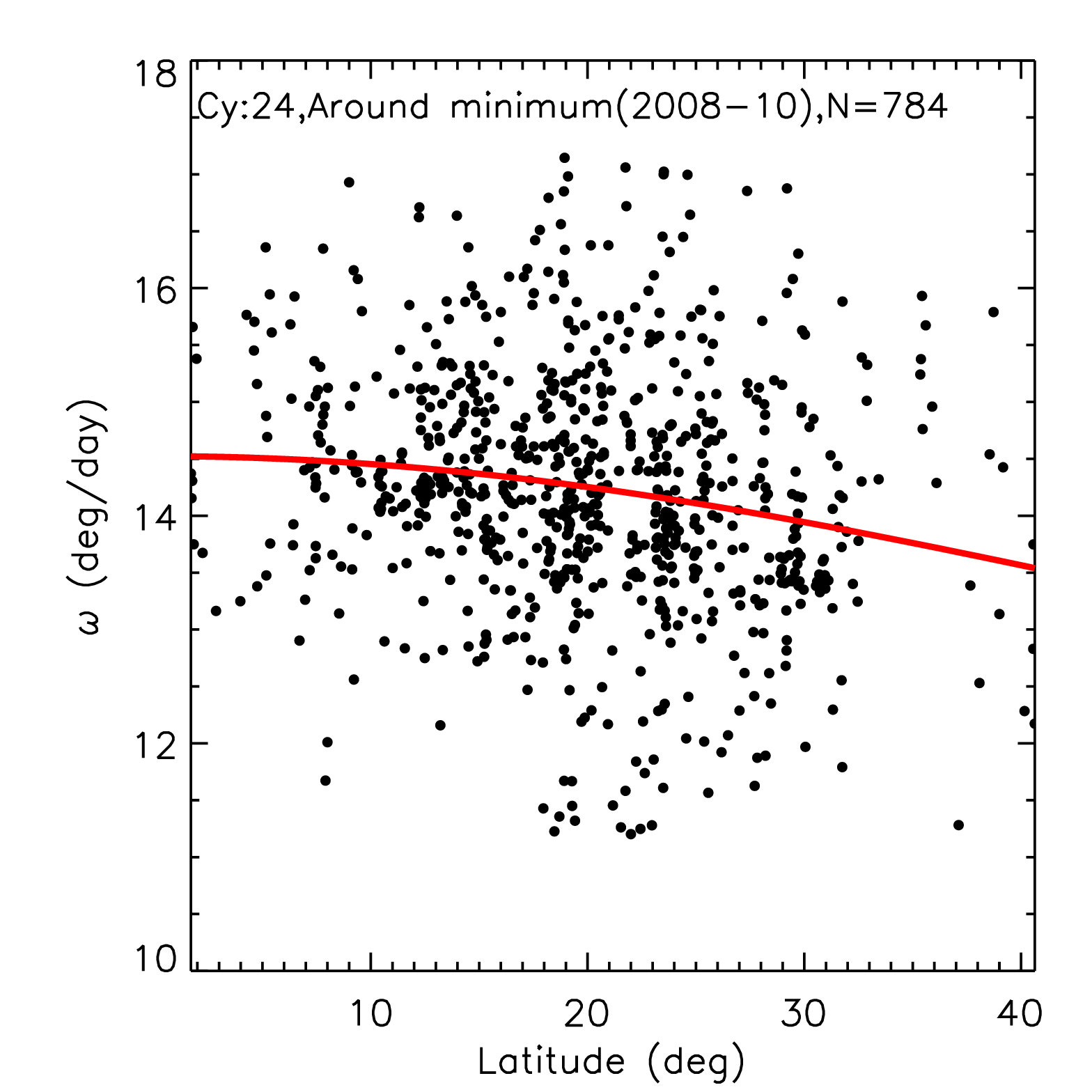}
\end{subfigure}
\setcounter{figure}{2}
\caption{Latitude dependence of angular velocity ($\omega$) 
 determined from the daily data of sunspot groups around the minima
(left panels)  and around the maxima ({\it right panels}) of Solar Cycles 23 
({\it top panels}) and 24 ({\it middle panels}). The {\it bottom panel} 
represents  the latitude dependence of $\omega$ in  interval 2008--2010.
Northern- and southern-hemispheres' data are combined.
 The {\it continuous curve} (red) represents the  profile of the differential
 rotation deduced from the corresponding  values of the coefficients 
$A$ and $B$ given in Table~2.}.
\label{f3}
\end{figure}

\begin{figure}
\setcounter{figure}{1}
\centering
\begin{subfigure}
\setcounter{figure}{0}
\includegraphics[width=5.6cm]{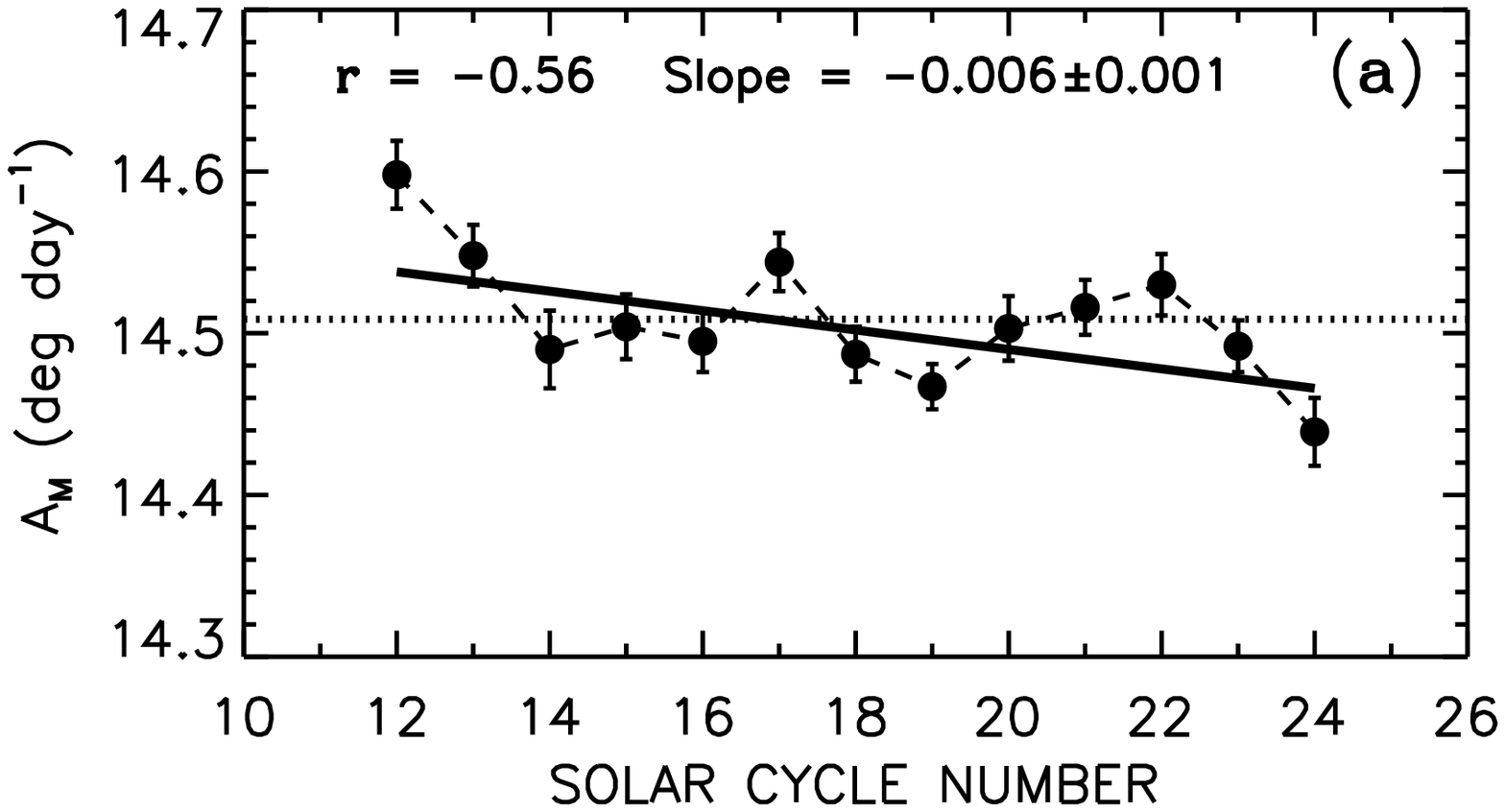}
\includegraphics[width=5.6cm]{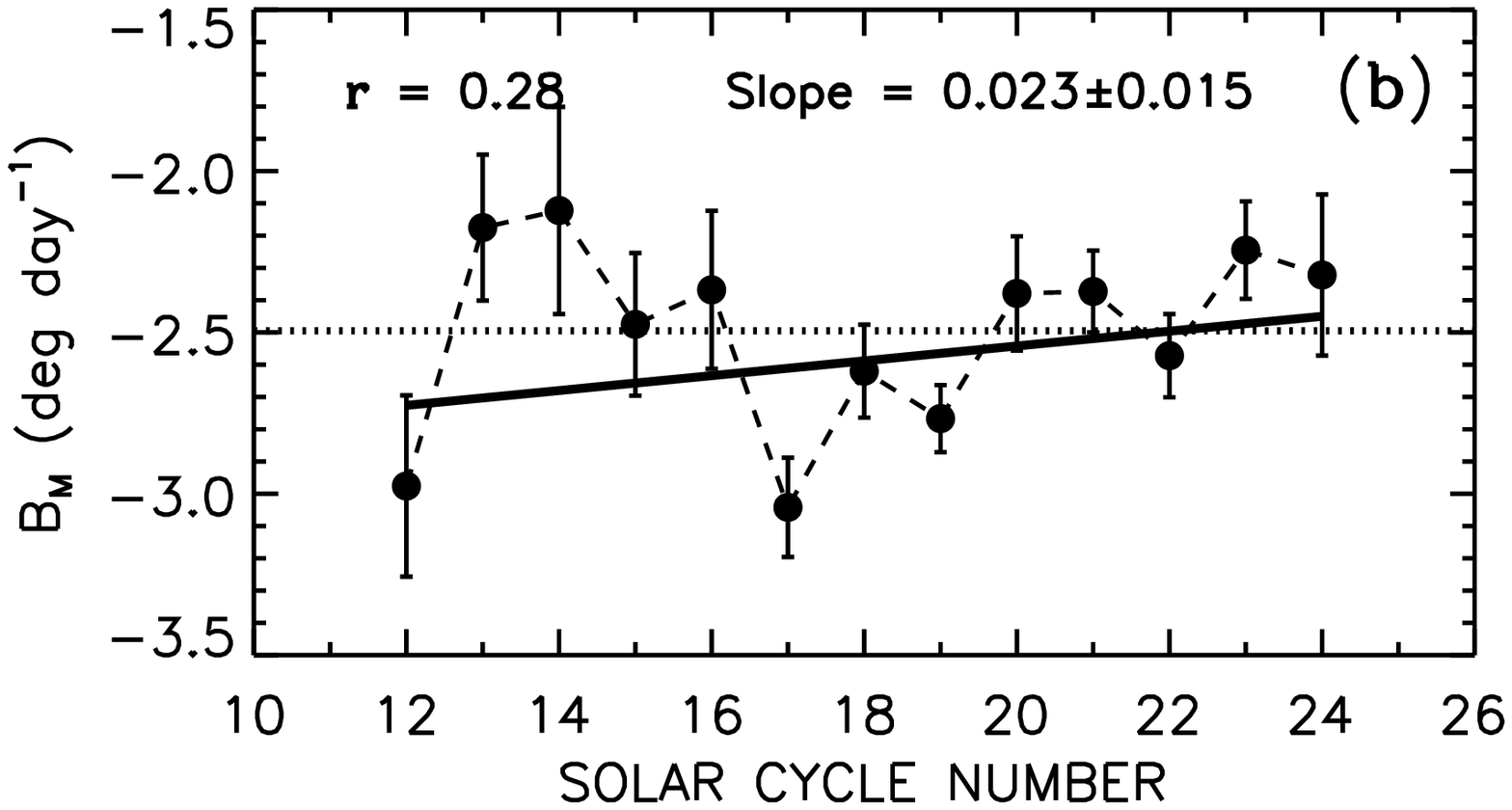}
\end{subfigure}
\begin{subfigure}
\setcounter{figure}{4}
\includegraphics[width=5.6cm]{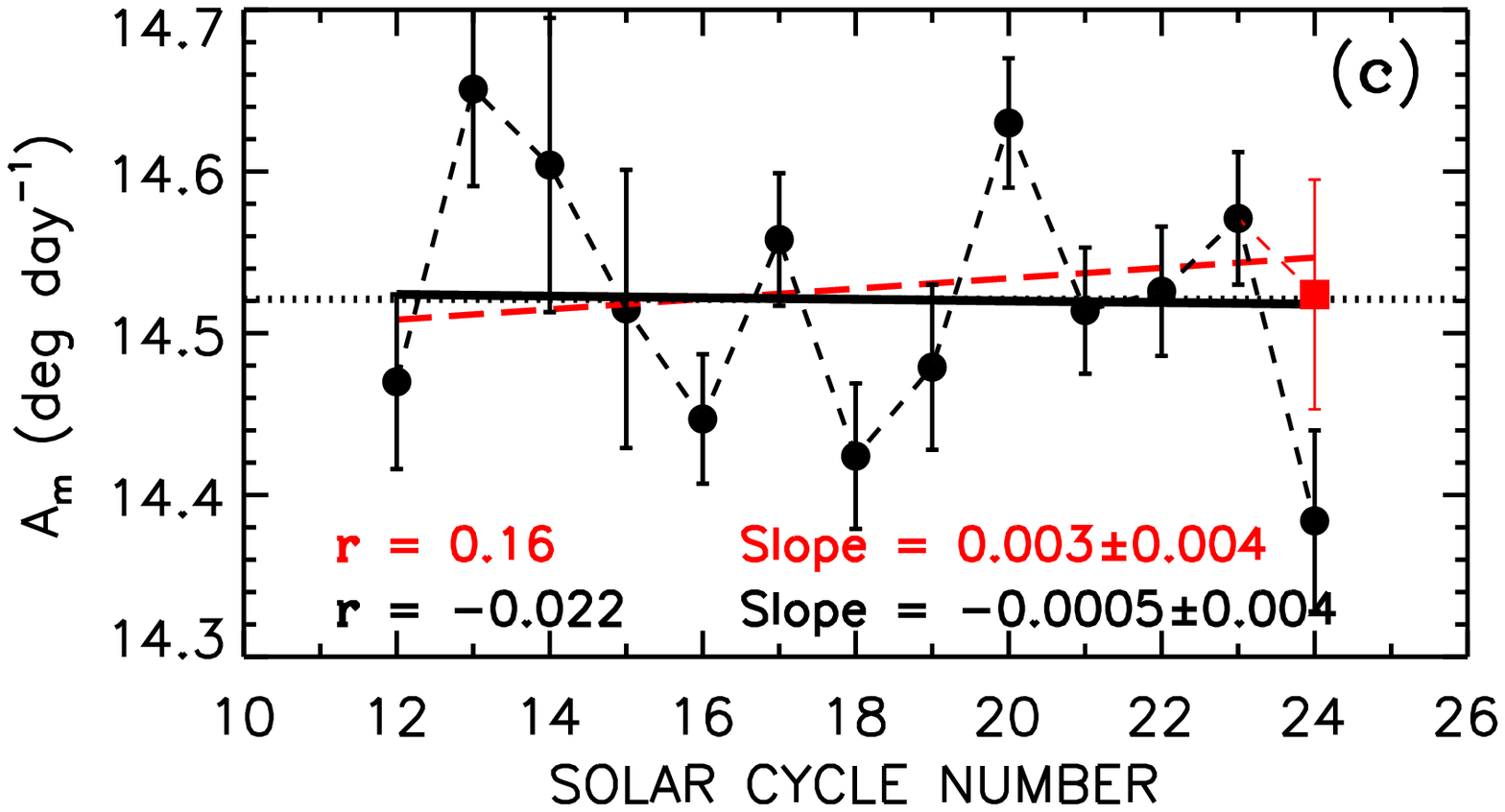}
\includegraphics[width=5.6cm]{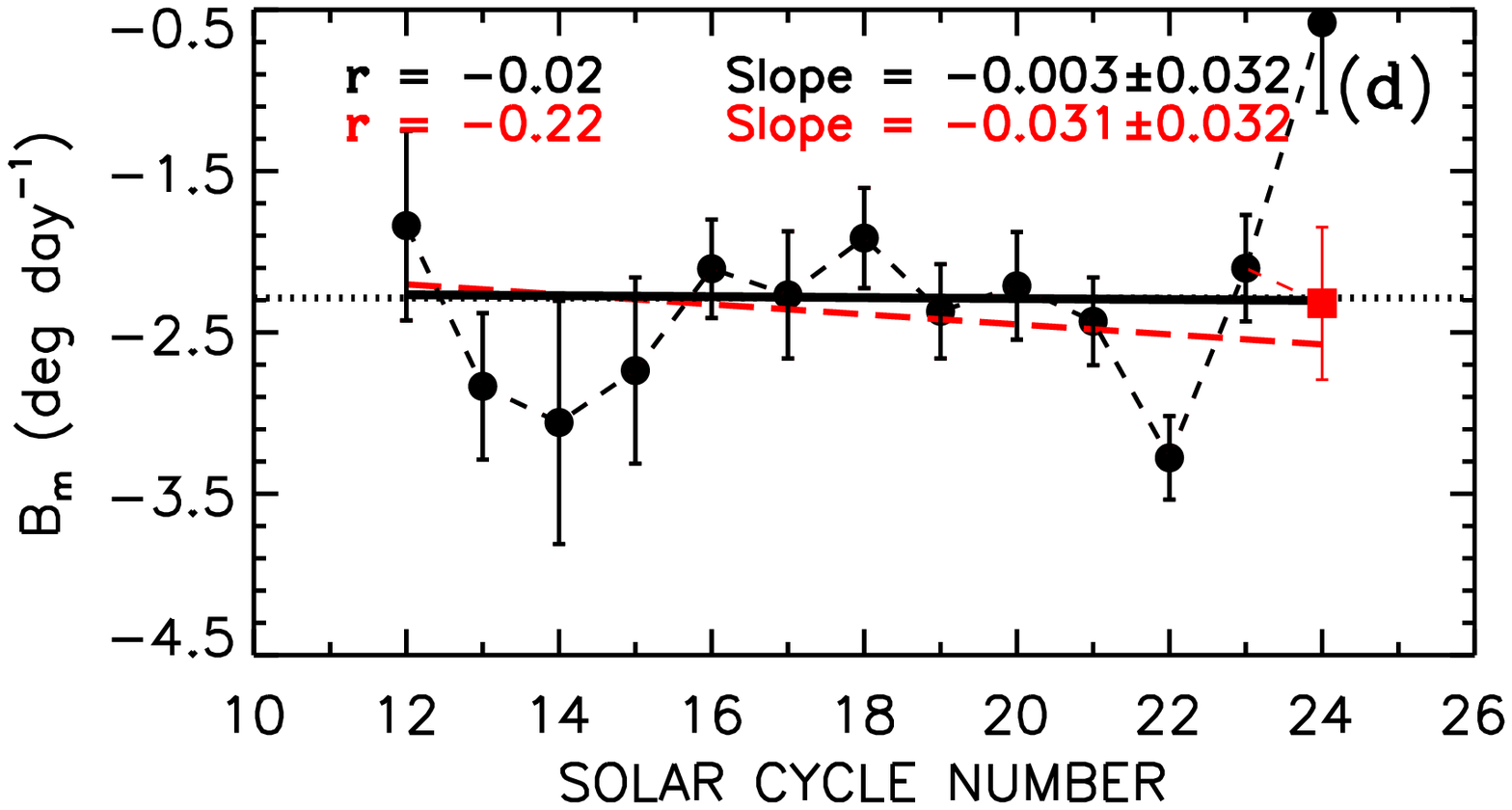}
\end{subfigure}
\caption{The {\it filled circle-dashed curves} represents  
the cycle-to-cycle variations  in the equatorial rotation
 rates [$A_{\rm M}$] and [$A_{\rm m}$], and the latitude gradients
 [$B_{\rm M}$] and [$B_{\rm m}$],  of solar rotation determined
from the sunspot-group data around the maxima and minima
(indicated by the suffixes  M and m, respectively) of  Solar Cycles 
 12\,--\,24. The values of $A_{\rm m}$ and  $B_{\rm m}$ in interval
 2008--2010 are  also shown ({\it red-filled-square}).    
 Northern- and southern-hemispheres' data 
are combined. The  horizontal {\it dotted-line} represents the mean.  
The {\it continuous line} represents the linear best-fit 
({\it red-long-dashed line} represent 
the best-fit when 2009 is considered for minimum epoch of Cycle~24). 
 The values of the slope and the  correlation coefficient are also shown.}  
\label{f4}
\end{figure}

Figure~2 shows variations in  $A$ and $B$ determined from the
 sunspot group
data  (northern- and southern-hemispheres' data are combined) 
in the 3-year MTIs 1875\,--\,1877, 1876\,--\,1878, $\dots$, 2015\,--\,2017
 during the period 1874\,--\,2017. 
In Table~2 we have given the values of $A$ and $B$ 
 around the maxima and minima
 of  Solar Cycles 12\,--\,24 obtained from the aforementioned 
3-year MTIs. These values are indicated in Figure~2 by the filled-circles and 
the filled-squares, respectively.  We have also shown the values of 
$A$ and $B$ obtained from the data in the interval 2008--2010. The reason  
for this  is given below.  
 The description of the yearly variations of 
$A$ and $B$ can be found in our  earlier papers, \inlinecite{jg95} and 
\inlinecite{jk99}, and in the paper by \inlinecite{ruz17}. Here we 
concentrated on cycle-to-cycle modulations in $A$ and $B$. 
\inlinecite{ruz17}  have noted
that when a significant peak of equatorial rotation velocity is observed during
minimum of activity, the strength of the next maximum is smaller than the
previous one. This is not found from  Figure~2.      
 
 Equation~1 is found to be not fitted well to the data in around minima 
of cycles. This is, obviously, because of insufficient data caused by large
 scarcity of sunspots at high latitudes of sunspot belt.
 As can be seen in Figure~2  
around the minimum of Solar Cycle~24  the values of $A$ and $B$ are 
abnormally low.  
 In Figure~3  we show
the latitude dependence of angular velocity during the 3-year intervals that 
comprise the epochs of  minima and the maxima of Solar Cycles~23 
and 24. 
 Northern- and southern-hemispheres' data are combined.
 In this figure the continuous curves represent the
corresponding profiles of the differential rotation deduced from the values
 of the coefficients $A$ and $B$ in Equation~1, which are also given in Table~2.
 As can be seen in this figure, the coefficients $A$ and $B$ reasonably 
well determined, $i.e.$  Equation~1 is reasonably well fitted to the data
 around the maxima of Solar Cycles~23 and 24 and even around the minimum of
 Solar Cycle~23. The corresponding fit is not good  around the minimum of
  Solar Cycle~24, $i.e.$ in the interval 2007--2009, because
 there is a large scarcity of data in higher latitudes of the sunspot 
latitude-belt (less overlap of Solar Cycles 23 and 24). Therefore, we have also 
checked the results  by considering tentatively  2009 for  
 minimum epoch of Solar Cycle~24 and  
in Figure~3 we have also shown the latitude dependence 
of the velocity in the interval 2008--2010 (middle year is 2009).
The fit of Equation~1 to the data of this interval  is reasonably good 
(Note: in the 13-month smoothed monthly 
mean of the revised SN,  2008.958, $i.e.$ around 15 November 2008, is the 
minimum epoch of Solar Cycle~24.)     

Figure~4 shows 
the cycle-to-cycle variations in the values of  
$A_{\rm M}$, $A_{\rm m}$, $B_{\rm M}$, and $B_{\rm m}$, 
 $i.e$ the values of  $A$ and $B$   around 
the maxima and  around the minima of Solar Cycles 12--24 
 given in Table~2
(the suffixes  M and m indicate epochs of sunspot maximum and minimum,  
 $i.e.$ the middle years of the 3-year intervals that comprise these 
years, respectively). For the reason given above, besides the values 
of $A$ and $B$ in the interval 2007--2009 that comprises 2008 in its middle, 
we have also shown the corresponding  values in the interval 2008--2010 
that comprises  2009 in its middle.
 The linear least-square fit correspond to each of these parameters is  
done by taking uncertainty in the  parameter, $i.e.$ 
${\rm weight} = \frac{1}{\sigma^2}$ is used. 
   In Article-I the existence of a secular decreasing trend was found 
in the cycle-to-cycle modulation of $A$ determined from the 
 sunspot group data (over the whole cycle) 
 of  Solar Cycles 12\,--\,22, but it was not found in the corresponding  
modulation in $B$.  
As can be seen in  Figure~4 
there exist no  notable secular trends in both  $A_{\rm m}$ and 
  $B_{\rm m}$, whereas   
there is a considerably significant secular decreasing-trend in
  $A_{\rm M}$ (the slope is about five times larger than  
its standard deviation).  The decreasing trend of  $A_{\rm M}$
is slightly less steeper (slope $-0.006 \pm 0.001$) than that of 
the average $A$ over the whole cycle (slope $-0.01 \pm 0.001$) found in 
Article-I.  There is a secular  decreasing trend in $B_{\rm M}$
($i.e.$ the secular decreasing trend in the latitudinal gradient around
 maxima of solar cycles), but it is not well defined (the slope is only 
about 1.5 times of  its standard deviation). 
 We find the following correlations ($r$ represents the 
correlation coefficient  and the values given within parentheses
 are found when 2009 is considered for the minimum epoch of Solar Cycle~24):\\
 $r = 0.35$ ($0.11$) between $A_{\rm m}$ and $A_{\rm M}$, \\
 $r = 0.33$ ($0.48$) between $A_{\rm m}$ and $B_{\rm M}$, \\
 $r = -0.65$ ($-0.46$)  between $A_{\rm m}$ and $B_{\rm m}$, \\
 $r = -0.45$ between $A_{\rm M}$ and $B_{\rm M}$, \\
 $r = -0.36$ ($0.02$) between $A_{\rm M}$ and $B_{\rm m}$, and \\
 $r = -0.12$ ($-0.39$) between $B_{\rm m}$ and $B_{\rm M}$. \\
 The anticorrelation 
between $A$ and $B$ implies that an increase in the  equatorial rotation rate  
with an increase in the latitude gradient of rotation, and vise-versa 
(it should be noted that  $B$ is negative). 

We find no significant correlation between 
sunspot number (either the value of sunspot minimum [$R_{\rm m}$] or  
of maximum [$R_{\rm M}$])   
and $A$ or $B$ (the highest correlation, 
$r = -0.39$,  is found between $B_{\rm M}$  and $R_{\rm M}$).
However, the existence of secular decrease of $A$ with secular 
increase of solar activity seems to be some extent established 
(see \opencite{li14}; \opencite{os16}; \opencite{ruz17}, 
 and the references therein). 
 The existence of a positive correlation between 
$A$ determined from the whole-cycle data and length of cycle is known 
(\opencite{mend99}). In Article-I a relatively weak correlation  
  ($r = 0.57$) was found between $A$ and cycle-length 
from the data of  Solar Cycles 12\,--\,22 and based on that  
 a short length was predicted for Cycle~23. 
 That  prediction is incorrect (Cycle~23 is  a long  cycle).
Here the correlation between 
$A_{\rm M}$ and cycle-length is found  highly insignificant. 
The following  reasonably large  correlations ($r > 0.6$) are
 found  (the values given within parentheses are found when 
2009 is considered for the minimum epoch of Solar Cycle~24):\\
 $r = 0.67$ ($0.67$) between $A_{\rm m}$ and cycle length,\\
 $r = 0.77$ ($0.77$) between $A_{\rm m}$ and cycle declining time, and\\  
 $r = 0.64$ ($0.44$) between $B_{\rm m}$ and cycle rising time.\\

Using the value of  $A_{\rm m}$ of Solar Cycle~24 in  the linear
 relationship correspond to the aforementioned highest
 correlation between $A_{\rm m}$ and cycle-declining time (determined 
from the data of Cycles~12\,--\,23), we obtained $5.69\pm 0.42$ years for the 
declining time of Cycle~24, which  suggests that Solar Cycle~24 will end at 
$2019.98 \pm 0.42$, $i.e.$ within the interval
 July 2019\,--\,May 2020. This may be close to reality. 
 However, the aforementioned  correlations may not strong enough for 
prediction purpose. 

Due to the large equatorial rotation rate
there could be a strong Coriolis force at the minimum of a solar cycle, which
greatly affects (decreases) the emergence of toroidal magnetic flux at
low latitudes (\opencite{chg87}). Depending on the strength of
 $A_{\rm m}$ this effect might persist throughout the corresponding 
solar cycle.
In fact,  a reasonable correlation is also exist
 between $A_{\rm m}$ and cycle-length. 
 Coriolis forces  act
 more effectively on small magnetic features, that may reduce the rate of 
emergence of toroidal flux  at low latitudes
 (the effect of magnetic 
buoyancy force is large on large magnetic structures and  plasma drag force  
is large on small magnetic structures).  
  Hence, there is a reasonably  
good  correlation 
between  $A_{\rm m}$ and the declining time of the corresponding cycle, 
suggesting that a large  $A_{\rm m}$ might 
cause a long declining time. 
It should be noted that a large absolute value of $B$ means a large 
latitude gradient, $i.e.$ more differential rotation (strong dynamo).
The existence of a reasonably good correlation between  $B_{\rm m}$ and rising 
time of a cycle may imply that more differential rotation during 
the minimum of a cycle might cause the cycle to rise fast. This in turn related
 to the amount of toroidal magnetic flux production depending on the
 strength of the differential  rotation during the  minimum of the cycle. 
However, the physical relationships of all the aforementioned  correlations 
need  investigations.

\begin{figure}
\setcounter{figure}{2}
\centering
\begin{subfigure}
\setcounter{figure}{0}
\includegraphics[width=5.6cm]{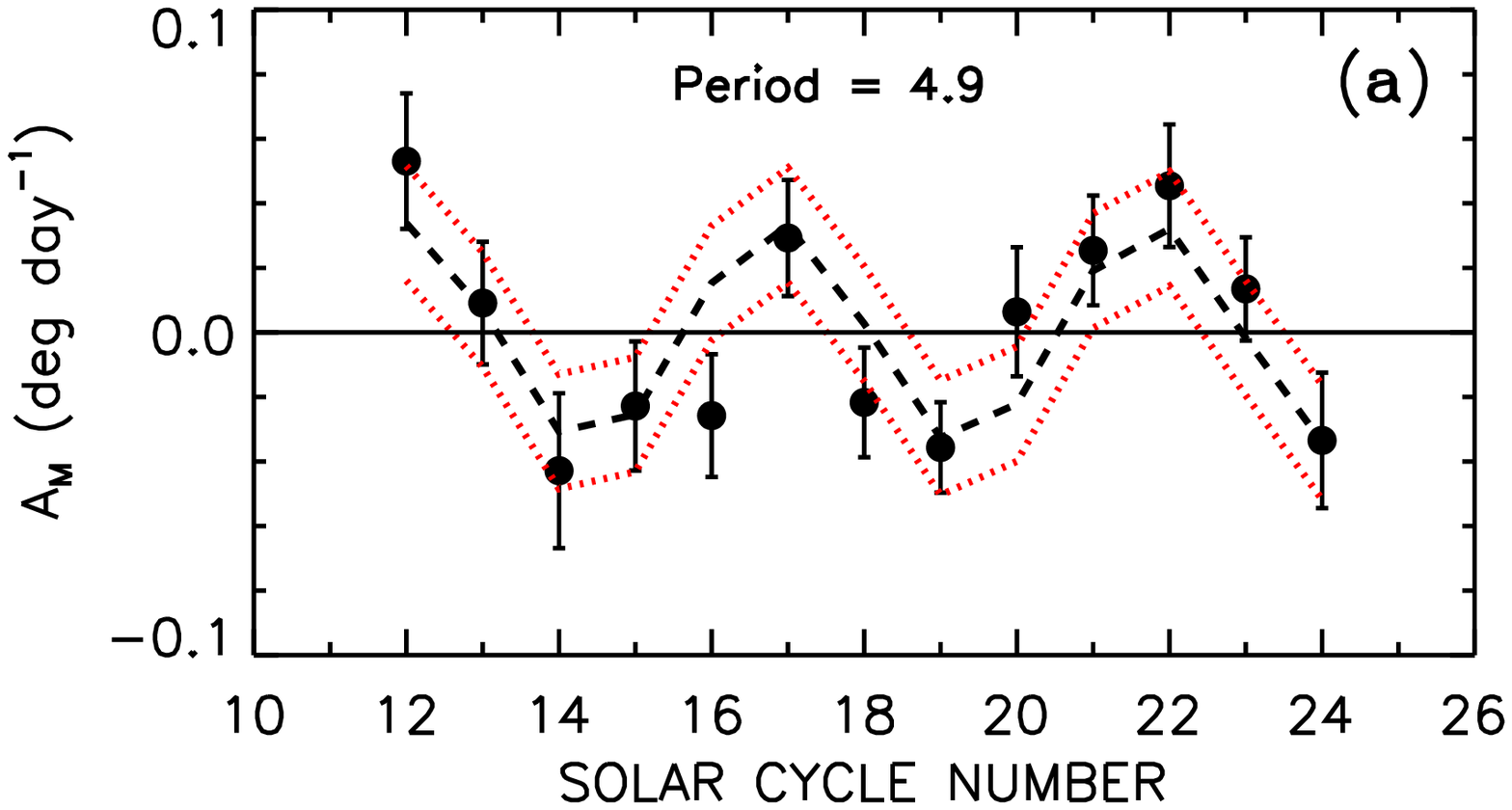}
\includegraphics[width=5.6cm]{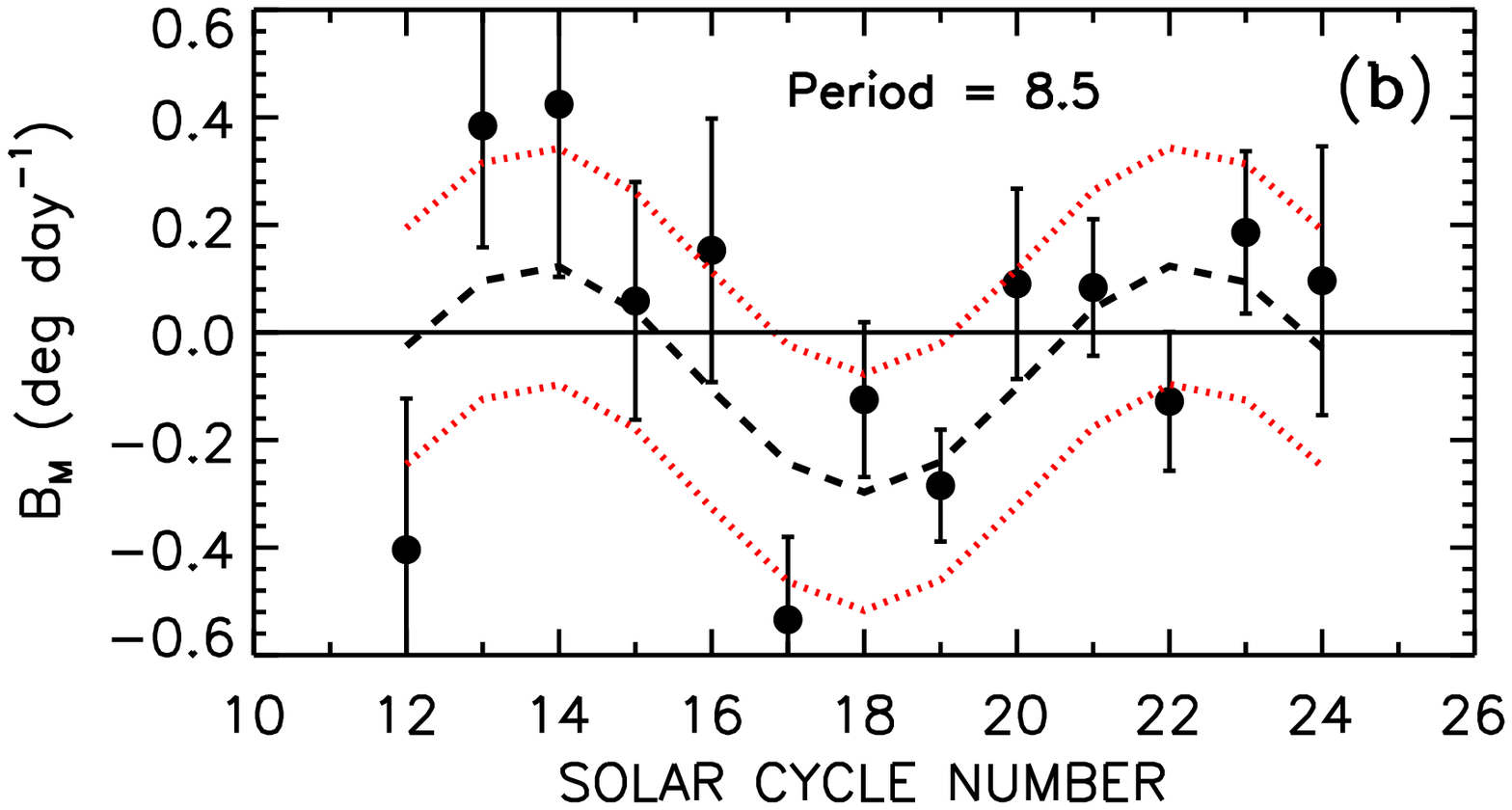}
\end{subfigure}
\begin{subfigure}
\setcounter{figure}{5}
\includegraphics[width=5.6cm]{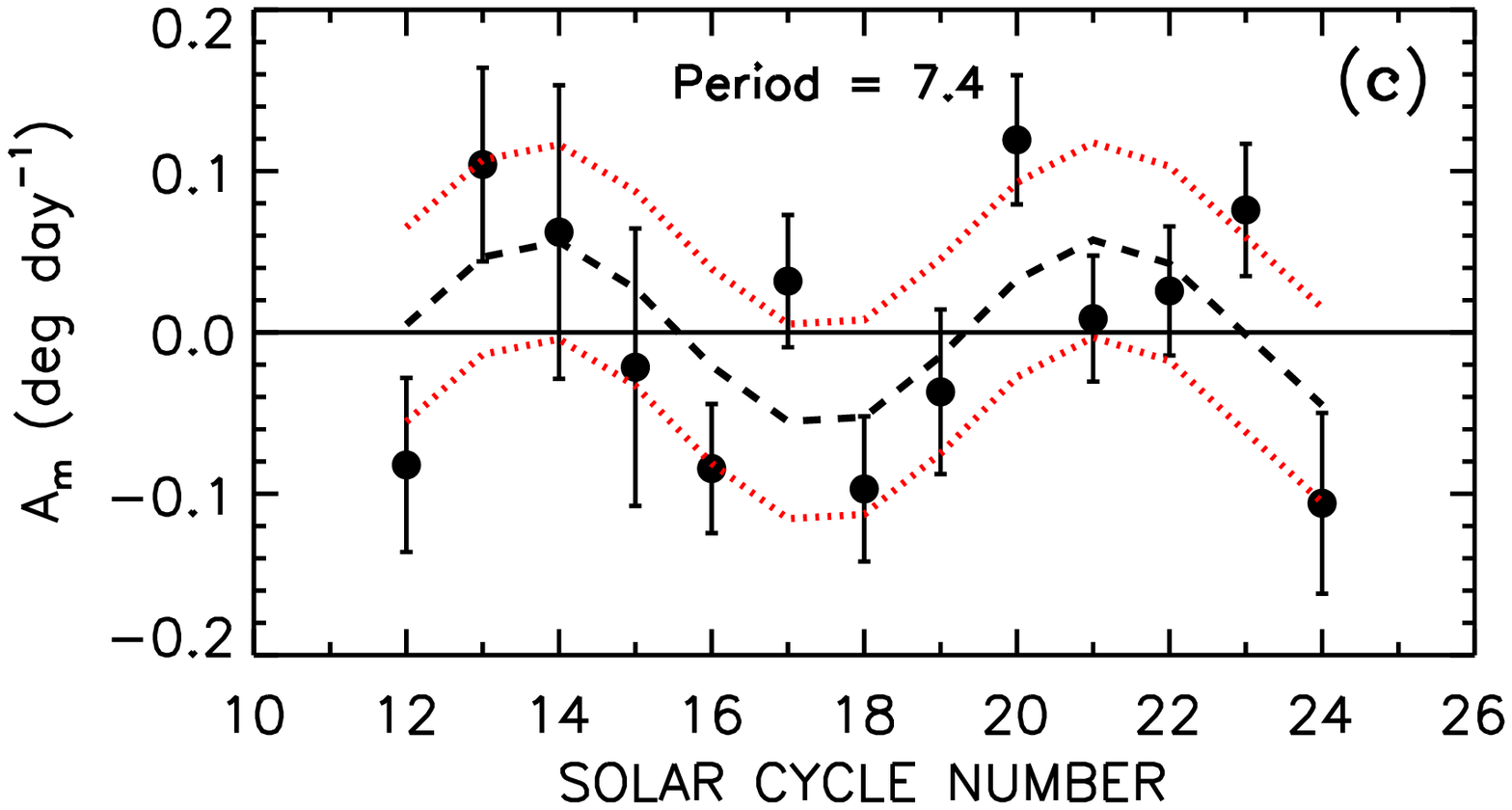}
\includegraphics[width=5.6cm]{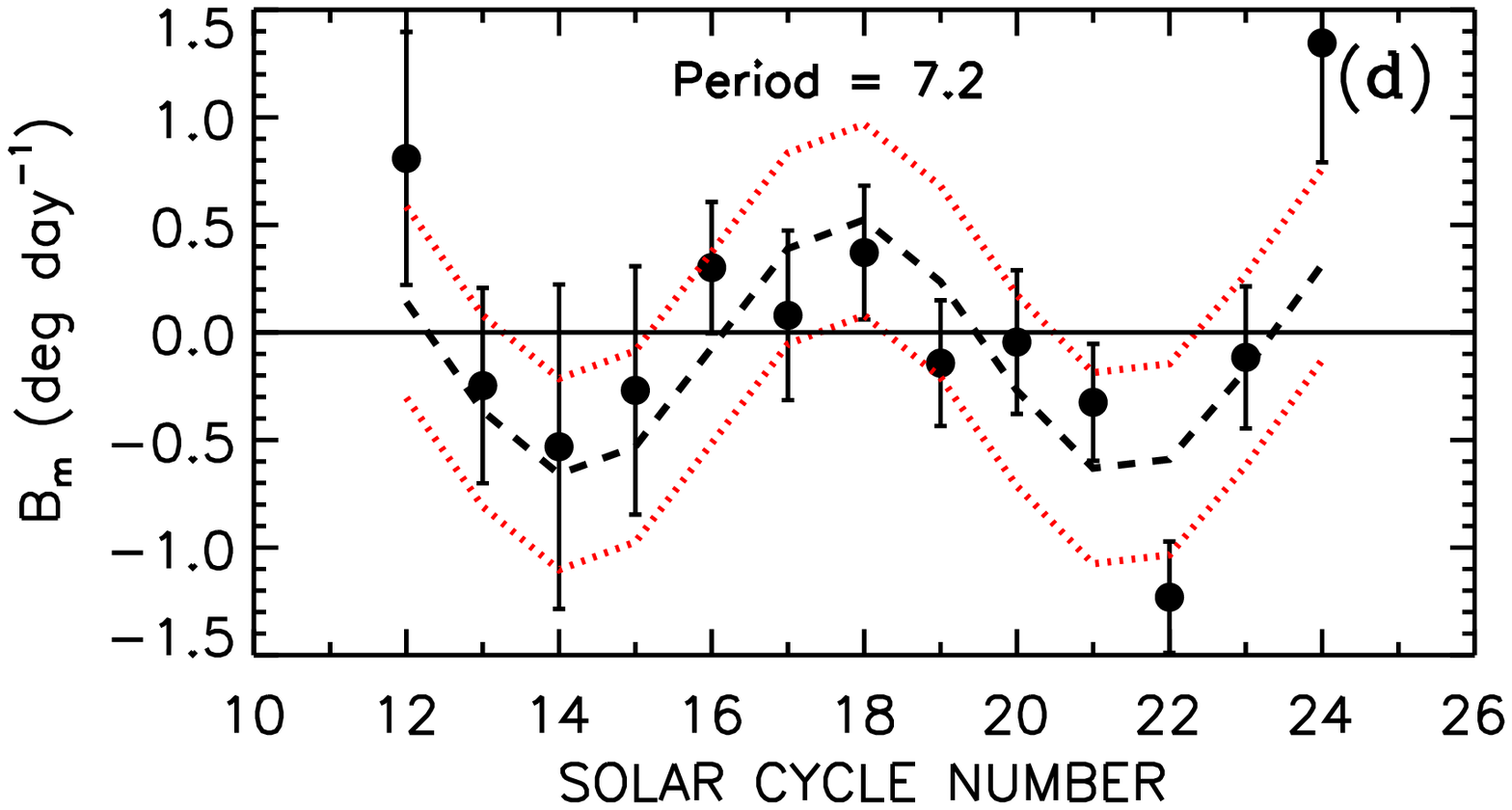}
\end{subfigure}
\caption{The {\it dashed curve} represents the best-fit cosine function
to the values  ({\it filled-circles}) of 
  ({\bf a}) $A_{\rm M}$, ({\bf b}) $B_{\rm M}$, ({\bf c}) $A_{\rm m}$, and 
({\bf d})  $B_{\rm m}$, 
after subtracting the corresponding  linear-fits. The corresponding values
of {\it root-mean-square-deviation} ({\it rmsd}) are 0.018, 0.22, 0.06, 
 and 0.44, respectively, and the 
the corresponding values of $\chi^2$ are 11.53, 16.35, 23.22, and 16.95,
 respectively (21.026 is the  5\,\% level significant value of $\chi^2$).    
The obtained periods of  $A_{\rm M}$, $B_{\rm M}$,  $A_{\rm m}$, and 
$B_{\rm m}$ are 4.9, 8.5, 7.4 and 7.2 (in number of solar cycles),  
 $i.e.$ $54.05 \pm 5.58$-year, $93.75 \pm 10.03$-year, $81.62 \pm 8.73$-year,
and $79.42 \pm 8.5$-year, respectively (Note: the average period of  solar
 cycles is $11.03 \pm 1.18$-year; see Pesnell, 2018.) 
 The {\it dotted curve} (red) represents the one-{\it rmsd}  level.}
\label{f5}
\end{figure}

\begin{figure}
\setcounter{figure}{4}
\centering
\begin{subfigure}
\setcounter{figure}{6}
\includegraphics[width=5.6cm]{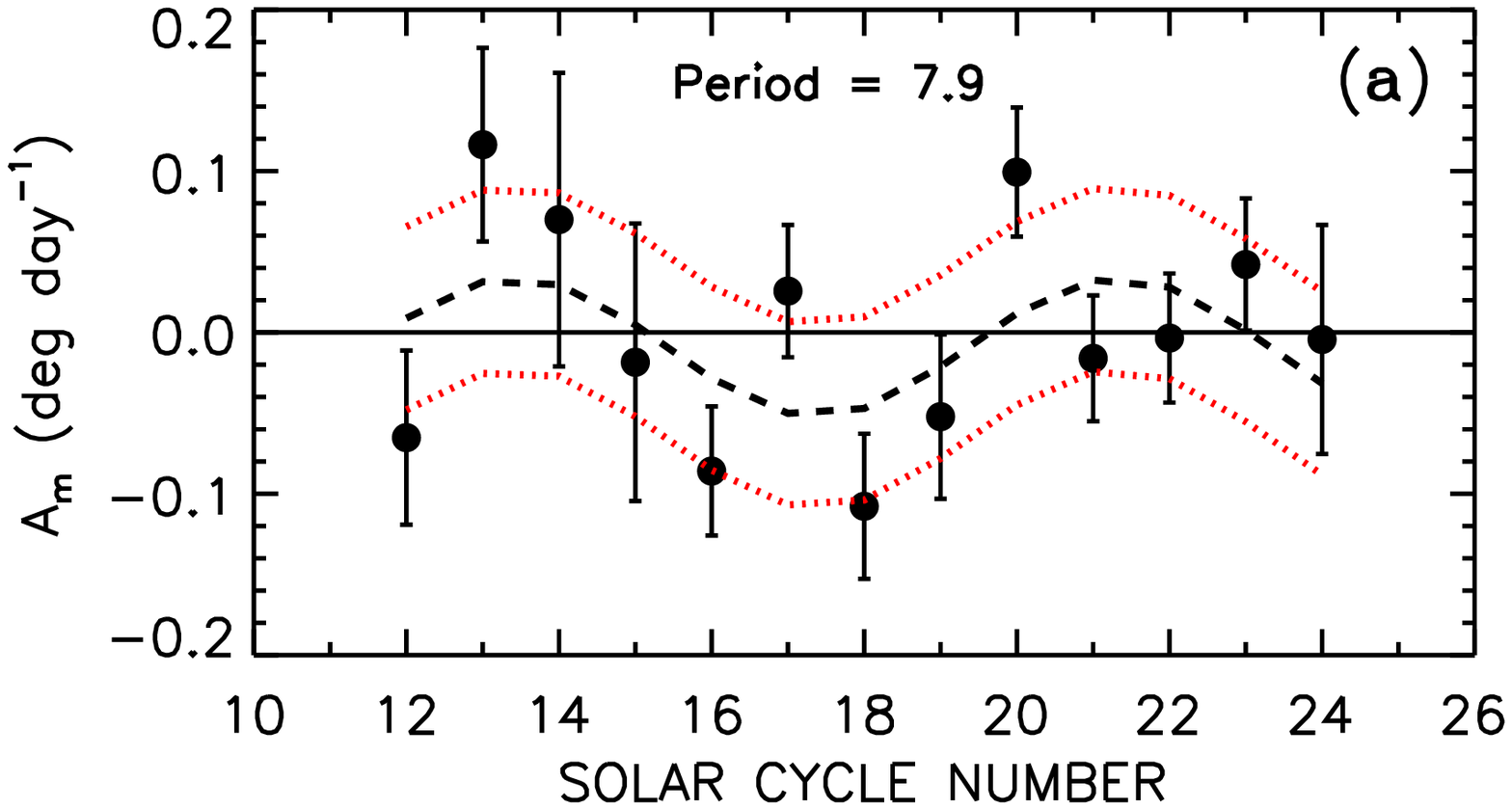}
\includegraphics[width=5.6cm]{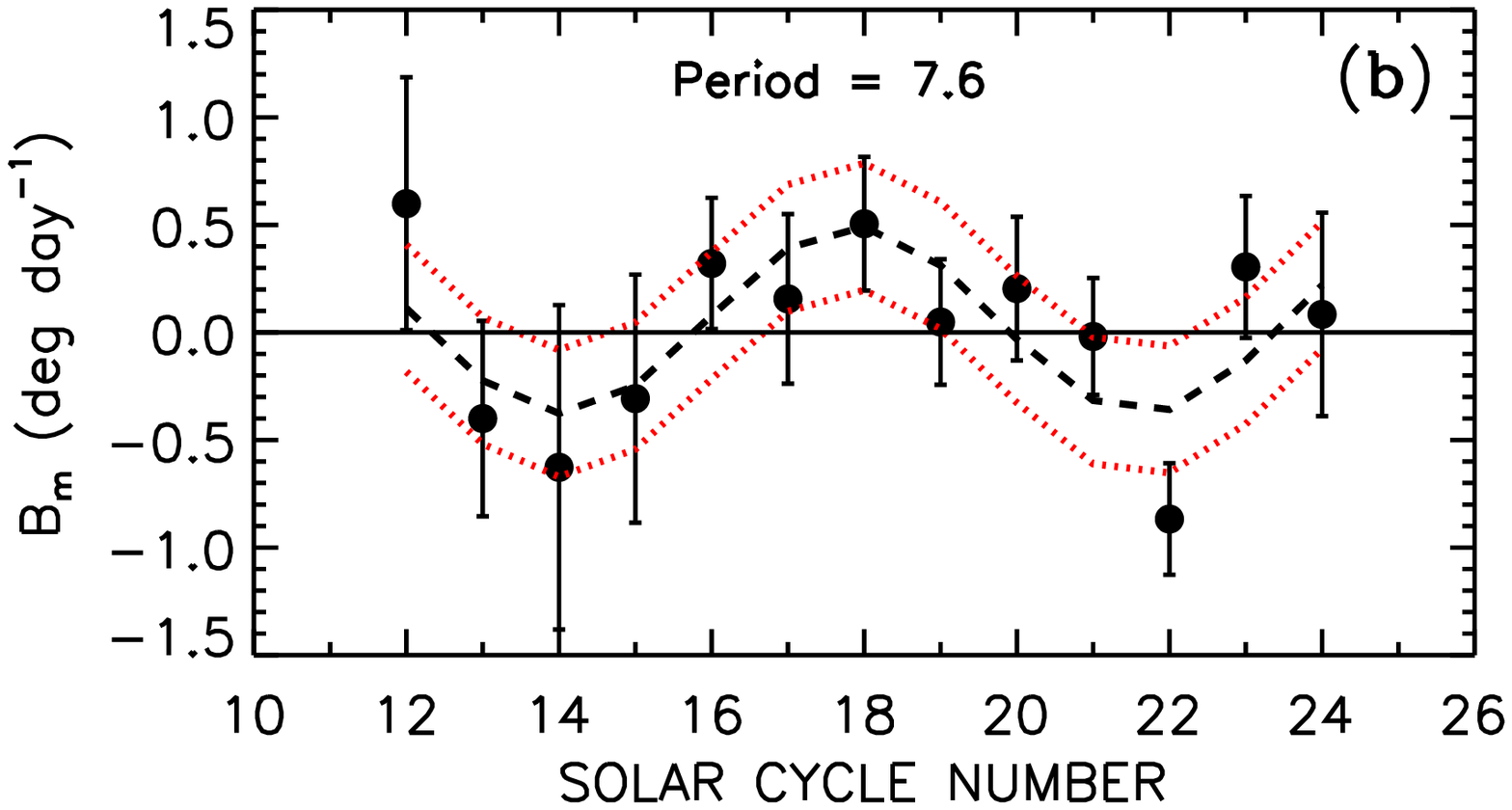}
\end{subfigure}
\caption{The {\it dashed curve} represents the best-fit cosine function
to the values  ({\it filled-circles}) of ({\bf a}) $A_{\rm m}$ and
 ({\bf b}) $B_{\rm m}$, by replacing the values of these parameters in
 interval 2007--2009 with that in interval 2008--2010.  The corresponding 
values of {\it root-mean-square-deviation} ({\it rmsd}) are 0.06  and 0.29, 
respectively, and the corresponding values of $\chi^2$ are 19.9 and 10.14
 respectively.  The obtained periods of  $A_{\rm m}$ and $B_{\rm m}$ are 
 7.9 and 7.6 (in number of solar cycles),   $i.e.$  $87.14 \pm 9.32$-year 
 and $83.83 \pm 9.0$-year, respectively. The {\it dotted curve} (red) 
represents the one-{\it rmsd}  level.}
\label{f6}
\end{figure}

Figure~5 shows  the best-fits of cosine functions 
to the values of
   $A_{\rm M}$,  $B_{\rm M}$,  $A_{\rm m}$, and   $B_{\rm m}$ that 
are given in Table~2,
after subtracting  the corresponding linear trends (in the cosine-fits  
${\rm weight} = \frac{1}{\sigma^2}$ is used).  
The cosine-fit to the data of most of the above parameters is reasonably 
good. Except in the case of the cosine-fit of $A_{\rm m}$, 
the $\chi^2$ of the cosine-fit to the data of each of the 
 remaining  parameters is reasonably small. In each of these cases, 
most of the data points are
  within the one-{\it rmsd}  level. 
The obtained periods of  $A_{\rm M}$, $B_{\rm M}$,  $A_{\rm m}$, and
$B_{\rm m}$ are 4.9, 8.5, 7.4 and 7.2 (in number of solar cycles),
$i.e.$ $54.05 \pm 5.58$-year, $93.75 \pm 10.03$-year, $81.62 \pm 8.73$-year,
and $79.42 \pm 8.5$-year, respectively (Note: the average period of a solar
 cycle is $11.03 \pm 1.18$-year; see \opencite{pesnell18}.)
  Considerable differences exist in the
 periodicities of  $A$,  and also in those of  $B$, that  
 correspond to the epochs of maxima and minima of solar cycles.
 The amplitude ($0.058^\circ$ day$^{-1}$) of the cosine profile of 
 $A_{\rm m}$ is  about 41\,\% larger than that ($0.034^\circ$ day$^{-1}$)
 of  $A_{\rm M}$, whereas the amplitude ($0.6^\circ$ day$^{-1}$) of 
the cosine profile of $B_{\rm m}$ is about 
65\,\% larger than that ($0.21^\circ$ day$^{-1}$) of $B_{\rm M}$.  
The cosine profiles of $A_{\rm M}$ and  $A_{\rm m}$ suggest the
 existence of a large   (up to 180$^\circ$)
 phase (initial) difference (approximate anticorrelation) between
 $A_{\rm M}$ and  $A_{\rm m}$.
The cosine profiles of $B_{\rm M}$ and  $B_{\rm m}$ also suggest the 
 existence of a large   (up to 180$^\circ$)
 phase (initial) difference (approximate anticorrelation) between
 $B_{\rm M}$ and  $B_{\rm m}$. The cosine profiles of $A_{\rm m}$ and  
$B_{\rm M}$ seem to be  approximately in phase (a positive 
correlation). 
The  cosine profile of $A_{\rm M}$ indicates that in the case   around
 maxima of Solar Cycles~12--24, the equatorial rotation rate
is maximum at Cycles~12, 17, and 22, and minimum at 
Solar Cycles 14, 19, and 24.
The  cosine profile of $A_{\rm m}$ indicates that
in the case of  around minima of Solar Cycles~12--24 
the equatorial rotation rate
is maximum at Solar Cycles~13/14 and 21, and minimum at 
Solar Cycles 17/18 and 24.
The  cosine profile of $B_{\rm M}$ indicates that 
in the case   around maxima of 
Solar Cycles~12--24, the latitude gradient of rotation 
is maximum at Solar Cycle~18 and minimum at Solar Cycles 14 and 22.
The  cosine profile of $B_{\rm m}$ indicates  that 
in the case of around minima of Solar Cycles~12--24
the latitude gradient of rotation
is maximum at Solar Cycles~14  and 21/22, and minimum at 
Solar Cycle~18 (it should be noted that $B_{\rm M}$ and  
$B_{\rm m}$ are having negative sign, a large negative value implies
more  differential rotation).

 For the reason given above, we have  repeated the calculations of 
 the cosine fits for $A_{\rm m}$ and $B_{\rm m}$ by replacing the values of  
these parameters in interval 2007--2009 with that in interval 2008--2010, 
$i.e.$ by considering the 2009, instead of 2008, for the minimum epoch
 of Solar Cycle~24. Figure~6a and 6b show the corresponding best-fits of cosine
 functions, which are more accurate than the 
corresponding ones shown in Figures~5c and 5b. That is, in the case of the  
cosine fits  of both $A_{\rm m}$ and $B_{\rm m}$  that are shown in 
Figures~6a and 6b the values of $\chi^2$  are smaller, and in the case of 
$B_{\rm m}$ the value of rmsd  is also smaller than the corresponding 
fits shown in Figures~5c and 5d. The cosine fits shown in Figure~6 indicate 
 the periodicities $87.14 \pm 9.32$-year and $83.83 \pm 9.0$-year 
for $A_{\rm m}$ and $B_{\rm m}$, respectively, which are slightly 
longer than  those indicated by the corresponding fits shown in 
Figure~5c and 5d.  Except these  all the suggestions made above  remain 
the same.

Cent percent anticorrelation 
between $A$ and $B$ implies  variations in  equatorial rotation rate 
and  magnitude  of latitude gradient are perfectly in phase. 
 Because of a large difference between the periods of $A_{\rm M}$ 
and $B_{\rm M}$, there could be a large  collapse (catastrophic) 
 in the corresponding phases during Solar  Cycles 12\,--\,24. 
The relatively less difference between the 
periods of $A_{\rm m}$ and $B_{\rm m}$  and the 
existence of an anticorrelation between the profiles of these parameters 
may suggest that  around the minimum of a solar cycle the  
  angular momentum  transport from higher latitudes to the equator may be 
more efficient. 
The cosine profiles  of $A_{\rm m}$ and $A_{\rm M}$ 
look to be in phase and in anti-phase during the alternate intervals of 
about 40\,--\,50-years (there exist a very weak correlation between  
$A_{\rm m}$ and $A_{\rm M}$; see above), causing reduction and improvement of 
flux emergence in the respective alternate intervals.  Since as mentioned 
above  the effect of $A_{\rm m}$ persists throughout the cycle, there exist
 an approximate coherence in  the cosine  profiles of $A_{\rm m}$ and 
  $B_{\rm M}$, which  my have a role in the toroidal flux production 
rate causing  variations in the amplitudes of Gleissberg cycles.

\section{Conclusions and Discussion}
Studies of variations in the solar differential rotation are important for
understanding the underlying mechanism of solar cycle and also other 
variations of solar activity. We analyzed GPR and DPD 
daily sunspot group data during the period 1874\,--\,1917
 and determined  the equatorial rotation rate [$A$] and the latitude
 gradient [$B$] of the  solar differential rotation by fitting the 
data in each 3-year MTI  during the period 1874\,--\,1917 
to the standard law of differential rotation.
The values of $A$ and $B$ around the years of maxima and
minima of  Solar Cycles 12\,--\,24   are obtained from the 3-year MTIs
series of $A$ and $B$ and studied
the long-term cycle-to-cycle modulations in these coefficients.
Here we have used the epochs of the maxima and  minima
 of  the  Solar Cycles 12\,--\,24 that
 were recently determined from the revised Version-2 international 
sunspot-number (SN) series. We  find that
there exits  a considerably significant secular decreasing-trend in
  $A$  around the maxima of solar cycles. There exist
 no secular trends in both  $A$ and $B$   around the minima of solar
cycles. The secular trend in $B$  around the  maxima
 of solar cycles is also found to be statistically insignificant.
  We fitted a cosine function to the values of $A$, and also
to those  of $B$, after removing the corresponding linear trends. 
The cosine-fits suggest
that there exist $\approx$54-year ($\approx$94-year) and 
$\approx$82-year ($\approx$79-year) periodicities in
$A$ ($B$)   around the maxima  and minima of solar cycles, 
respectively. The amplitude  of the cosine-profile of
 $A$ ($B$)   around the minima of solar cycles 
 is  about 41\,\% (65\,\%) larger than that 
 of  $A$ ($B$) of the  maxima. 
 In addition, the cosine profiles of $A$ and $B$ suggest  a 
large (up to $180^\circ$) phase difference between  the long-term variations
 of $A$, and also  between those of $B$,   around the maxima and 
minima of solar cycles.

 In Article-I the combined Greenwich and SOON sunspot-group
 data during Solar  Cycles 12--23 were analyzed and  shown the existence 
of a $7.4 \pm 0.5$-cycle in  $B$ by fitting a cosine-fit
to the cycle-to-cycle modulation of $B$. The similar periodic modulations 
were also seen in the values of $B$
 determined from the sunspot-group data of both the northern and
 the southern hemispheres (\opencite{jbu05b}). In 
\inlinecite{jbu05b} the cycle-to-cycle modulations in whole cycles $A$ and $B$
determined in Article-I were compared with the  modulations found in
 $A$ and $B$ determined using the 
 Kodaikanal Observatory 
(Solar Cycles: 15--21),  Mt. Wilson Observatory (Solar Cycles: 16--21), 
Solar-Observatory Kanzelh\"ohe (Solar Cycles: 18--21),  
 and National Observatory of Japan (Solar Cycles: 19-21) sunspot data bases 
 and pointed out the 
differences/similarities in the results  and some plausible causes for
 the differences.
\inlinecite{suz14} analyzed the
 sunspot-group data measured in Mt. Wilson Observatory and  their own
 measurements during  Solar Cycles~16--23 and found the existence of a  
6--7-cycle in the cycle-to-cycle
 modulation of $B$ (absolute values) determined from  the whole-sphere data  
and also in the modulations of  $B$ determined from the 
northern- and southern-hemispheres' data. The 
  pattern of the modulation of $B$ during Solar Cycles~16--23 that 
is found by \inlinecite{suz14} closely similar to 
 the corresponding portion  (from Cycle~16 to Cycle~23) of the
 modulation (of $|B|$) during Solar  Cycles 12--23
 found in Article-I, as well as to that of $B$ around maxima of the 
 cycles in the current analysis. Also, it seems to closely fit to the 
corresponding portion of the 
cosine-curve of $B$ of  whole cycles  in Article-I and also to that of
 $B$  around the maxima of the cycles in the present analysis. 
Overall, all these results seem to strongly confirm 
the existence of about a 7-cycle modulation in $B$ determined from 
sunspot data. In \inlinecite{suz14} it is found that   in Solar Cycle~16 
the northern hemisphere value of $B$ is significantly smaller than that of 
the southern-hemisphere and  almost monotonic decrease in the magnitude of 
$B$ in both the hemispheres from Solar Cycles-17 to 21. The overall 
pattern seems
 to be suggesting  the existence of a phase-shift (occurred in Cycle~16)  
between the  northern and  southern hemispheres.  In \inlinecite{jbu05b}
 it is found that  except in Cycle~12 the magnitude of the
  northern-hemisphere $B$ is highly significantly larger than that of the 
southern hemisphere, in each of the remaining 
cycles the difference  between  
the northern- and southern-hemispheres' values of $B$ is statistically 
insignificant. That is,  no phase difference is found between 
the cycle-to-cycle modulations in $B$ of northern and southern hemispheres. 
In the present analysis we have not studied the north--south asymmetry
 of the modulations in $A$ and $B$ around  the maxima and the  minima 
of solar cycles. In \inlinecite{suz14}
 error bars of $B$ values are relatively small. However, surprisingly  
in the largest Cycle~19 the error bar of $B$ seems to be largest. 
 Particularly,  the error bar of the northern-hemisphere's
  value of $B$ of this cycle is much larger than  that of the 
southern hemisphere and  in fact, larger than those of both the 
 northern- and southern-hemisphere's values of the remaining all cycles. 
In Cycle~19 activity in northern hemisphere was  larger that 
of southern hemisphere ($e.g.$ \opencite{tem06}; \opencite{jj19}). 
 The gaps in  Greenwich data due to   missing days of
 observations are filled with the  data from  other observatories. Such an
 update of Mt. Wilson sunspot-group data may be necessary.
\inlinecite{suz14}  found that
 the modulation in $B$   determined from  the data around the 
middle years (unequal intervals)
 of the solar cycles is almost the same as that of $B$ 
determined from the data of whole cycles. The arguments above on 
the  modulation in $B$ of whole cycles also apply to  
 the modulation in $B$   determined from  the data around the 
middle years of the cycles. 
The modulations  in $A$ and $B$   around the maxima of cycles 
found here are also closely similar to those   
 of $A$ and $B$  determined from the data of the whole cycles  shown in 
Article-I. This could be  due to a large  portion of the data of a 
 cycle are coming from around the maximum  of the cycle, $i.e.$ the ratio
 of the size of a whole-cycle's data to that of the data around the maximum is
 considerably small. In many studies of solar cycle-variations of $A$ 
and $B$  it has been 
found that the magnitudes of $A$ and $B$ are much larger, and larger
fluctuations in the yearly variations of $A$ and $B$,  
during the minima of many solar cycles than during the maxima 
($e.g$. \opencite{balth86}; \opencite{jg95}, \opencite{jj03}).  
 \inlinecite{suz14}  found the same in the yearly variations of $B$
and also found that if the value of $B$ at minimum/maximum of a cycle 
is large, then the cycle-average value of $B$ is large. In spite of 
we have used here the 3-year intervals, there exist considerably large 
fluctuations in the values of $A$ and $B$ around the  minima of 
some cycles (see Figure~2).
We find the existence of  
an anticorrelation between the cycle-to-cycle modulations of $B$  around the 
maxima and the minima of cycles and 
for the  reason mentioned above  
the modulation in $B$ around the 
maxima is almost the same as that of $B$ of the whole cycle.
 The average  value of $B$   over all cycles maxima  is found 
slightly larger that of over all cycles minima.
 However, the amplitude of the cosine-fit of
 $B$ around the minima of cycles  is much larger than that of $B$ 
around the maxima.

The $\approx$54-year periodicity in 
$A$   around the maxima of solar cycles  found here might confirm the 
$\approx$55-year
 grand cycle of $A$ was found  by \inlinecite{yk93} using
National Astronomical Observatory of Japan (NAOJ)  sunspot-group 
data during the period 1948\,--\,1987. The known  $\approx$55-year periodicity 
of sunspot activity  ($e.g.$ \opencite{tan11}; \opencite{gao16}; 
\opencite{komit16}) may be related to the $\approx$54-year period of $A$. 
The well-known Gleissberg cycle of solar activity may be related to 
the $\approx$94-year period  in $B$   around maxima of solar cycles   and 
 to  both the 82-year and the 79-year periods in $A$ and $B$, respectively,
    around  minima of solar cycles. That is, causes of the  periodicities 
of solar activity may be have influences of the periodicities 
in $A$ and $B$ (also see Article-I; \opencite{jb16}). 
However, what causes  the variations  in $A$ and $B$ is not known yet, except 
that there are  speculations for the existence of the Sun's spin--orbit 
coupling (\opencite{wood65}; \opencite{juc03}; \opencite{jj05}).  

The  maximum and minimum  epochs of a solar cycle comprise 
relatively large and small numbers of large sunspot groups, respectively. 
The magnetic structures of large and small sunspot groups may anchor  at 
relatively  deep and shallow layers of the solar convection zone, respectively.
 Hence, the periodicities in $A$ and $B$  around the maxima and
 minima of solar cycles are different and might  originate at relatively 
deep and shallow layers of the solar convection zone, respectively. However, 
the observed rotation rate of a sunspot group (in general a magnetic tracer) 
does not need to be the rotation rate of plasma at the anchoring depth of 
the sunspot
 group. The difference between the plasma and sunspot-group rotation rates 
 depends upon the effects of the dynamical forces like magnetic buoyancy, 
Coriolis force, and drag force  on the magnetic structure of the sunspot
 group  ($e.g.$ \opencite{dh94}).  Therefore,  the cause of 
 the differences  in the periodicities of $A$ and $B$ 
  around the   maxima and minima of solar cycles still needs to be 
established.

\section{Acknowledgments}
The author thanks the anonymous referee for  the critical review and 
for  useful comments and suggestions. 
The author acknowledges the work of all the
 people contribute  and maintain the GPR and DPD  sunspot databases.
The sunspot-number data are provided by WDC-SILSO, Royal Observatory of
Belgium, Brussels. 

\section{Conflict of interest} 
The author declares to have no conflicts of interest.

{}
\end{article}

\begin{thebibliography}{}
\bibitem[\protect\citeauthoryear{Babcock}{1961}]{bab61}
Babcock, H.W.: 1961, {\it Astrophys. J.}  {\bf 133}, 572.
 DOI: \doiurl{10.1086/14706}
\bibitem[\protect\citeauthoryear{Badalyan and Obridko}{2017}]{bo17}
Badalyan, O.G., Obridko, V.N.: 2017, {\it Mon. Not. Roy. Astron. Soc.} 
{\bf 466}, 4535. DOI: \doiurl{10.1093/mnras/stx134}
\bibitem[\protect\citeauthoryear{Balthasar, V\'azquez, and W\"ohl}{1986}]{balth86}
Balthasar, H.,  V\'azquez, M., W\"ohl, H.: 1986,  {\it Astron. Astrophys.} 
{\bf 155}, 87. ADS: http://adsabs.harvard.edu/abs/1986A\%26A...155...87B
\bibitem[\protect\citeauthoryear{Baranyi, Gy\H{o}ri, Ludm\'any}{2016}]{bara16}
Baranyi, T., Gy\H{o}ri, L., Ludm\'any, A.: 2016, {\it Solar Phys.}
 {\bf 291}, 3081. DOI: \doiurl{10.1007/s11207-016-0930-1}
\bibitem[\protect\citeauthoryear{Braj\v{s}a, Ru\v{z}djak, and W\"ohl}{2006}]{braj06}
Braj\v{s}a, R.,  Ru\v{z}djak, D., W\"ohl, H.: 2006, {\it Solar Phys.} {\bf 237}, 365. DOI: \doiurl{10.1007/s11207-006-0076-7}
\bibitem[\protect\citeauthoryear{Cameron, Dikpati, and Brandenburg}{2017}]{camer17}
Cameron, R.H., Dikpati, M., Brandenburg, A.: 2017 {\it Space Sci. Rev.} 
{\bf 210}, 367. DOI: \doiurl{10.1007/s11214-015-0230-3}
\bibitem[\protect\citeauthoryear{Chandra, Vats, and Iyer}{2010}]{cvi10}
Chandra, S., Vats, H.O.,  Iyer, K.N.: 2010, {\it Mon. Not. Roy. Astron. Soc.}  {\bf 407}, 1108.
 DOI: \doiurl{10.1111/j.1365-2966.2010.16947.x}
\bibitem[\protect\citeauthoryear{Choudhuri and Gilman}{1987}]{chg87}
Choudhuri, A.R., Gilman, P.A.: 1987, {\it Astrophys. J.}, {\bf 316}, 788.
DOI: \doiurl{10.1086/165243}
\bibitem[\protect\citeauthoryear{Dikpati and Gilman}{2006}]{dg06}
Dikpati, M., Gilman, P.A.: 2006,  {\it Astrophys. J.} {\bf 649}, 498.
 DOI: \doiurl{10.1086/506314}
\bibitem[\protect\citeauthoryear{D'Sliva and Howard}{1994}]{dh94}
D'Silva, S., Howard, R.F.: 1994, {\it Solar Phys.} {\bf 151}, 213.
DOI: \doiurl{10.1007/BF00679072}
\bibitem[\protect\citeauthoryear{Gao}{2016}]{gao16}
Gao, P.X.: 2016, {\it Astrophys. J.}, {\bf 830}, 140.
DOI: \doiurl{10.3847/0004-637X/830/2/140}
\bibitem[\protect\citeauthoryear{Garcia and Mouradin}{1998}]{garc98}
Garcia, A., Mouradian, Z.: 1998, {\it Solar Phys.} {\bf 180}, 495.
DOI: \doiurl{10.1023/A:1005018712900}
\bibitem[\protect\citeauthoryear{Gilman and Howard}{1984}]{gh84}
Gilman, P.A., Howard, R.: 1984, {\it Astrophys. J.} {\bf 283}, 385.
 DOI: \doiurl{10.1086/162316}
\bibitem[\protect\citeauthoryear{Gupta, Sivaraman, and Howard}{1999}]{gupta99}
Gupta, S.S., Sivaraman, K.R.,  Howard, R.: 1999, {\it Solar Phys.} {\bf 188},
 225. DOI: \doiurl{10.1023/A:1005229124554}
\bibitem[\protect\citeauthoryear{Gy\H{o}ri, Baranyi, and Ludm\'any}{2010}]{gyr10}
Gy\H{o}ri, L., Baranyi, T., Ludm\'any, A.: 2010, in {\it Proc. Intern. Astron.
Union 6, Sympo. S273}, {\bf 2011}, 403.
 DOI: \doiurl{10.1017/s174392131101564X}
\bibitem[\protect\citeauthoryear{Gy\H{o}ri, Ludm\'any, and Baranyi}{2017}]{gyr17}
Gy\H{o}ri, L., Ludm\'any, A., Baranyi, T.: 2017, 
{\it Mon. Not. Roy. Astron. Soc.}  {\bf  465}, 1259. 
DOI: \doiurl{10.1093/mnras/stw2667}
\bibitem[\protect\citeauthoryear{Hathaway}{2015}]{hath15}
Hathaway, D.H.: 2015,  {\it Living Rev. Solar Phys.} {\bf 12}, N0.4. 
DOI:\doiurl{10.1007/lrsp-2015-4}
\bibitem[\protect\citeauthoryear{Hathaway, Wilson, and Reichmann}{1999}]{hath99}
Hathaway, D.H., Wilson, R.M., Reichmann, E.J.: 1999, {\it J. Geophys. Res.}
 {\bf 104}, 22375. DOI: \doiurl{10.1029/1999JA900313}
\bibitem[\protect\citeauthoryear{Hiremath}{2002}]{hi02}
Hiremath, K.M.: 2002, {\it Astron. Astrophys.} {\bf 386}, 674.
 DOI: \doiurl{10.1051/0004-6361:20020276}
\bibitem[\protect\citeauthoryear{Howard}{1976}]{hr76}
Howard, R.: 1976, {\it Astrophys. J. Lett.} {\bf 93}, L159.
 DOI: \doiurl{10.1086/182328}
\bibitem[\protect\citeauthoryear{Howard, Gilman, and Gilman}{1984}]{hgg84}
Howard, R., Gilman, P.I.,  Gilman, P.A.: 1984, {\it Astrophys. J.} {\bf 283},
 373. DOI: \doiurl{10.1086/162315}
\bibitem[\protect\citeauthoryear{Howard and LaBonte}{1980}]{hl80}
Howard, R., LaBonte, B. J.: 1980, {\it Astrophys. J.}, {\bf 239}, L33.
DOI: \doiurl{10.1086/183286}
\bibitem[\protect\citeauthoryear{Howe \etal}{2000}]{howe00}
Howe, R., Christensen-Dalsgaard, J., Hill, F., Komm, R.W., Larsen, R.M.,
Schou, J., Thompson, M.J.,  Toomre, J.: 2000,  {\it Astrophys. J.}  {\bf 533},
 L163. DOI: \doiurl{10.1086/312623}
\bibitem[\protect\citeauthoryear{Javaraiah}{2003}]{jj03}
Javaraiah, J.: 2003, {\it Solar Phys.}  {\bf 212}, 23.
 DOI: \doiurl{10.1023/A:1022912430585}
\bibitem[\protect\citeauthoryear{Javaraiah}{2005}]{jj05}
Javaraiah, J.: 2005, {\it Mon. Not. Roy. Astron. Soc.} {\bf 362}, 1311.
DOI: \doiurl{10.1111/j.1365-2966.2005.09403.x}
\bibitem[\protect\citeauthoryear{Javaraiah}{2011}]{jj11}
Javaraiah, J.: 2011, {\it Adv. Space Res.} {\bf 48}, 1032.
 DOI: \doiurl{10.1016/j.asr.2011.05.004}
\bibitem[\protect\citeauthoryear{Javaraiah}{2013}]{jj13}
Javaraiah, J.: 2013, {\it Solar Phys.}, {\bf 287}, 197.
 DOI: \doiurl{10.1007/s11207-013-0345-1}
\bibitem[\protect\citeauthoryear{Javaraiah}{2017}]{jj17}
Javaraiah, J.: 2017, {\it Solar Phys.} {\bf  292}, 172.
DOI: \doiurl{10.1007/s11207-017-1197-x}
\bibitem[\protect\citeauthoryear{Javaraiah}{2019}]{jj19}
Javaraiah, J.: 2019, {\it Solar Phys.} {\bf 294}, 64.
DOI: \doiurl{10.1007/s11207-019-442-6}
\bibitem[\protect\citeauthoryear{Javaraiah and Bertello}{2016}]{jb16}
Javaraiah, J., Bertello, L.: 2016, {\it Solar Phys.} {\bf 291}, 3485.
 DOI: \doiurl{10.1007/s11207-016-1003-1}
\bibitem[\protect\citeauthoryear{Javaraiah, Bertello, and Ulrich}{2005a}]{jbu05a}
Javaraiah, J., Bertello, L.,  Ulrich, R.K.: 2005a (Article-I), {\it Solar Phys.}
 {\bf 232}, 25.  DOI: \doiurl{10.1007/s11207-005-8776-y}
\bibitem[\protect\citeauthoryear{Javaraiah, Bertello, and Ulrich}{2005b}]
{jbu05b}
Javaraiah, J., Bertello, L.,  Ulrich, R.K:  2005b, {\it Astrophys. J.}
{\bf 626}, 579. DOI: \doiurl{10.1086/429898}.
\bibitem[\protect\citeauthoryear{Javaraiah and Gokhale}{1995}]{jg95}
Javaraiah, J.,  Gokhale, M.H.: 1995, {\it Solar Phys.}  {\bf 158}, 173.
 DOI: \doiurl{10.10.1007/BF00680841}
\bibitem[\protect\citeauthoryear{Javaraiah and Gokhale}{1997b}]{jg97}
Javaraiah, J., Gokhale, M.H.: 1997, {\it Astron. Astrophys.} {\bf 327}, 795.
\bibitem[\protect\citeauthoryear{Javaraiah and Komm}{1999}]{jk99}
Javaraiah, J., Komm, R.W.: 1999, {\it Solar Phys.}  {\bf 184}, 41.
DOI: \doiurl{10.1007/s11207-006-0130-5}
\bibitem[\protect\citeauthoryear{Javaraiah and Ulrich}{2006}]{ju06}
Javaraiah, J.,  Ulrich, R.K.:  2006, {\it Solar Phys.} {\bf 237}, 245.
 DOI: \doiurl{10.1007/s11207-006-0130-5} 
\bibitem[\protect\citeauthoryear{Javaraiah \etal}{2009}]{jub09}
Javaraiah, J., Ulrich, R.K.,  Bertello, L., Boyden, J.E.:  2009,
{\it Solar Phys.}  {\bf 257}, 61. DOI: \doiurl{10.1007/s11207-009-9342-9}
\bibitem[\protect\citeauthoryear{Juckett}{2003}]{juc03}
 Juckett, D.A.: 2003, {\it Astron. Astrophys.} {\bf 399}, 731.
 DOI:   \doiurl{10.1051/0004-6361:20021923}
 \bibitem[\protect\citeauthoryear{Kambry and Nishikawa}{1990}]{kn90}
Kambry, M.A.,  Nishikawa, J.: 1990, {\it Solar Phys.}  {\bf 126}, 89.
DOI: \doiurl{10.1007/BF00158300}
\bibitem[\protect\citeauthoryear{Khutsishvili, Gigolashvili, and Kvernadze}{2002}]{kgk02}
Khutsishvili, E.V., Gigolashvili, M.SH., Kvernadze, T.M.: 2002, 
{\it Solar Phys.}  {\bf 206}, 219. DOI: \doiurl{10.1023/A:1015068629350}
\bibitem[\protect\citeauthoryear{Komitov \etal}{2016}]{komit16}
Komitov, B., Sello, S., Duchlev, P.,  Dechev, M.,  Penev, K.,
Koleva, K.: 2016,  {\it Bulg. Astronomic. J.} {\bf 25}, 78.
\bibitem[\protect\citeauthoryear{Komm, Howard, and Harvey}{1993}]{khh93}
Komm, R.W, Howard, R.F., Harvey, J.W.: 1993, {\it Solar Phys.}, {\bf 143}, 19.
DOI: \doiurl{10.1007/BF00619094}
\bibitem[\protect\citeauthoryear{LaBonte and Howard}{1982}]{lh82}
LaBonte, B.J., Howard, R.: 1982, {\it Solar Phys.} {\bf 75}, 161.
 DOI: \doiurl{10.1007/BF00153469}
\bibitem[\protect\citeauthoryear{Li \etal}{2014}]{li14}
 Li, K.J., Feng, W., Shi, X.J.,  Xie, J.L.,  Gao, P.X., 
Liang, H.F: 2014, {\it Solar Phys.} {\bf 289}, 759.  
 DOI: \doiurl{10.1007/s11207-013-0369-6}
\bibitem[\protect\citeauthoryear{Makarov, Tlatov, and Callebaut}{1997}]{mak97}
Makarov, V.I., Tlatov, A.G., Callebaut, D.K.: 1997, {\it Solar Phys.} {\bf 170},
 373.
 DOI: \doiurl{10.1023/A.1004995826593}
\bibitem[\protect\citeauthoryear{Mendoza}{1999}]{mend99}
Mendoza, B.: 1999, {\it Solar Phys.} {\bf 188}, 237.
 DOI: \doiurl{10.1023/A:1005256728266}
\bibitem[\protect\citeauthoryear{Obridko and Shelting}{2016}]{os16}
Obridko, V.N., Shelting, B.D.: 2016, {\it Astron. Lett.} {\bf  42}, 631.
DOI: \doiurl{10.1134/S1063773716080041}
\bibitem[\protect\citeauthoryear{Ogurtsov \etal}{2002}]{ogurt02}
Ogurtsov, M.G., Nagovitsyn, Y.A., Kocharov, G.E., Jungner, H.: 2002,
{\it Solar Phys.} {\bf 211}, 371. DOI: \doiurl{10.1023/A:1022411209257}
\bibitem[\protect\citeauthoryear{Olemskoy and Kitchatinov}{2005}]{ok05}
Olemskoy, S.V.,  Kitchatinov, L.L.: 2005, {\it Astron. Lett.} {\bf 31}, 706.
DOI: \doiurl{10.1134/1.2075313}
\bibitem[\protect\citeauthoryear{Pesnell}{2018}]{pesnell18}
Pesnell, W.D.: 2018,  {\it Space Weather} {\bf 16}, 1997.
DOI: \doiurl{10.1029/2018SW002080}
\bibitem[\protect\citeauthoryear{Roudier \etal}{2018}]{roudi18}
 Roudier, Th., \v{S}vanda, M., Ballot, J., Malherbe, J.M., Rieutord, M.: 2018, 
{Astron. Astrophys.} {\bf 611}, A92. DOI: \doiurl{10.1051/0004-6361/201732014}
\bibitem[\protect\citeauthoryear{Rozelot}{1994}]{roze94}
Rozelot, J.P.: 1994, {\it Solar Phys.} {\bf 149}, 149.
 DOI: \doiurl{10.1007/BF00645186}
\bibitem[\protect\citeauthoryear{Ru\v{z}djak \etal}{2017}]{ruz17}
Ru\v{z}djak, D., Braj\v{s}a, R., Sudar, D., Skoki\'c, I., 
Poljan\v{c}i\'c-Beljan, I.: 2017, {\it Solar Phys.} {\bf 292}, 179. 
DOI \doiurl{10.1007/s11207-017-1199-8}  
\bibitem[\protect\citeauthoryear{Singh and Prabhu}{1985}]{jsb85}
Singh, J., Prabhu, T.P.: 1985, {\it Solar Phys.} {\bf 97}, 203.
 DOI: \doiurl{10.1007/BF00152989}
\bibitem[\protect\citeauthoryear{Sivaraman \etal}{2003}]{siva03}
Sivaraman, K.R., Sivaraman, H., Gupta, S.S.,  Howard, R.: 2003,
{\it Solar Phys.} {\bf 214}, 65.
DOI: \doiurl{10.1023/A:1024075100667}
\bibitem[\protect\citeauthoryear{Snodgrass}{1992}]{snod92}
Snodgrass, H.B.: 1992, In: Harvey, K.L.(ed.) The Solar Cycle,
 {\bf CS-27}, {\it Astron. Soc. Pac.}, San Francisco, 205.
\bibitem[\protect\citeauthoryear{Snodgrass and Howard}{1985}]{sh85}
Snodgrass, H.B., Howard, R.: 1985, {\it Solar Phys.} {\bf 95}, 221.
 DOI: \doiurl{10.1007/BF00152399}
\bibitem[\protect\citeauthoryear{Sudar \etal}{2014}]{sudar14}
Sudar, D., Skok\'c,  Ru\v{z}djak, D., Brajsa, R., W\"ohl, H.: 2014,
{\it Mon. Not. Roy. Astron. Soc.} {\bf 439}, 1471.
 DOI: \doiurl{10.1093/mnras/stu099}
\bibitem[\protect\citeauthoryear{Suzuki}{2014}]{suz14}
 Suzuki, M.: 2014, {\it Solar Phys.} {\bf 289}, 4021.
 DOI: \doiurl{10.1007/s11207-014-0576-9}
\bibitem[\protect\citeauthoryear{\v{S}vanda \etal}{2008}]{svand08}
\v{S}vanda, M., Klv\v{a}na, M., Sobotka, M., Bumba, V.: 2008,
{\it Astron. Astrophys.} {\bf 477}, 285. 
DOI: \doiurl{10.1051/0004-6361:20077718}
\bibitem[\protect\citeauthoryear{Tan}{2011}]{tan11}
Tan, B.: 2011, {\it Astrophys. Space Sci.}, {\bf 332}, 65.
 DOI: \doiurl{10.1007/s10509-010-0496-6}
 \bibitem[\protect\citeauthoryear{Temmer \etal}{2006}]{tem06}
Temmer, M., Ryb\'bak, J., Bend\'ik, Veronig, A., Vogler, F., Otruba, W., 
P\"otzi, W., Hanslmeier, A.: 2006, {\it Astron. Astrophys.} {\bf 447}, 735.
DOI: \doiurl{10.1051/0004-6361:20054060}
\bibitem[\protect\citeauthoryear{Ward}{1965}]{war65}
Ward, F.: 1965, {\it Astrophys. J.} {\bf 141}, 534.
DOI: \doiurl{10.1086/148143}
\bibitem[\protect\citeauthoryear{Ward}{1966}]{war66}
Ward, F.:  1966, {it Astrophys. J.} {\bf 145}, 416. 
 DOI: \doiurl{10.1086/148783}
\bibitem[\protect\citeauthoryear{Wood and Wood}{1965}]{wood65}
Wood, R.M., Wood, K.D.: 1965, {\it Nature} {\bf 208}, 129.
DOI: \doiurl{10.1038/208129a0}
\bibitem[\protect\citeauthoryear{Yoshimura and Kambry}{1993}]{yk93}
Yoshimura, H., Kambry, M.A.: 1993, {\it Solar Phys.}, {\bf 143}, 205. 
 DOI: \doiurl{10.1007/BF00646482}
\end{thebibliography}
\end{document}